\documentclass[12pt]{article}
\usepackage{pdproc,epsfig} 

\def\nubar{\overline{\nu}} 
\def\aprle{\buildrel < \over {_{\sim}}} 
\def\aprge{\buildrel > \over {_{\sim}}} 
\def\una{\frac{{\rm erg}}{{\rm s}\;{\rm Mpc}^3} }

\begin{document}
\renewcommand{\thefootnote}{\alph{footnote}}
  
\title{
PROBLEMS  IN HIGH ENERGY ASTROPHYSICS}

\author{PAOLO  LIPARI}

\address{INFN, sez. Roma ``La Sapienza'' \\
 {\rm E-mail: paolo.lipari@roma1.infn.it}}

\abstract{In this  contribution  we discuss 
some   of the main problems in high energy astrophysics,
and the perspectives to solve them 
using  different  types of  ``messengers'':
cosmic  rays, photons  and neutrinos.}
   
\normalsize\baselineskip=15pt


\section{Introduction}
The  birth  of high  energy astrophysics  can  be traced 
to nearly a  century ago, when the balloon 
flights of Victor  Hess established  that  a form  of
ionizing radiation, that  was soon given the name of  ``cosmic rays'',  
was   arriving from  outer space.
Soon it was  demonstrated  that these ``rays''  are
in fact ultrarelativistic  charged particles, mostly
protons  and fully ionized nuclei\footnote{
Electrons  have a  steeper  energy spectrum  and  constitute
a fraction  of only few percent of  the  flux;   small  quantities
of positrons  and  anti--protons  are also present.},
 with  a  spectrum 
that  extends to  extraordinary  high energies.
The existence of a large  flux of ultrarelativistic  particles
arrived  completely  unexpected, 
and its origin  remained  a  mistery  that only now  is
beginning  to be clarified.
The  main reason   for this   very long   delay  in developing
an  understanding of this  important physical
phenomenon is that
cosmic  rays (CR) do not point to their production  sites,  because
 their trajectory  is bent by the magnetic  fields   that permeates  both
interstellar and intergalactic space.

Today   we know  that  our universe contains  several  classes
of astrophysical  objects   where  non--thermal  processes
are  capable to accelerate  charged particles, 
both  leptons  ($e^\mp$)  or  hadrons (protons and nuclei),
to very high energy.  These   relativistic particles,
interacting inside or near  their  sources,  
can  produce   photons, and (in case of  hadrons)  
neutrinos  that  then  travel in  straight lines 
allowing  the imaging of the sources.
At  the  highest energy also the  magnetic  bending
of charged   particles can  become  sufficiently small to allow
source imaging.
A  detailed  study of the ``high energy universe''  can
therefore in principle be performed  using  
three   different   ``messengers'':
photons, neutrinos  and the cosmic ray themselves.
This  ``multi--messenger'' approach is still in its infancy,
but    has  the  potential to  give  us 
deep  insights about  the sites and 
physical  mechanisms  that  produce these very  high energy particles.

Many of  the  proposed  (or  detected)  acceleration 
sites are also associated with the violent acceleration of large 
macroscopic masses  (one example is  Gamma  Ray Bursts), and therefore
a  fourth ``messenger'':  gravitational waves   has the potential
to  give  very important and complementary information about these
astrophysical environments.

\section{The Cosmic Ray Spectrum}
\begin{figure} [hbt]
\centerline{\epsfig{figure=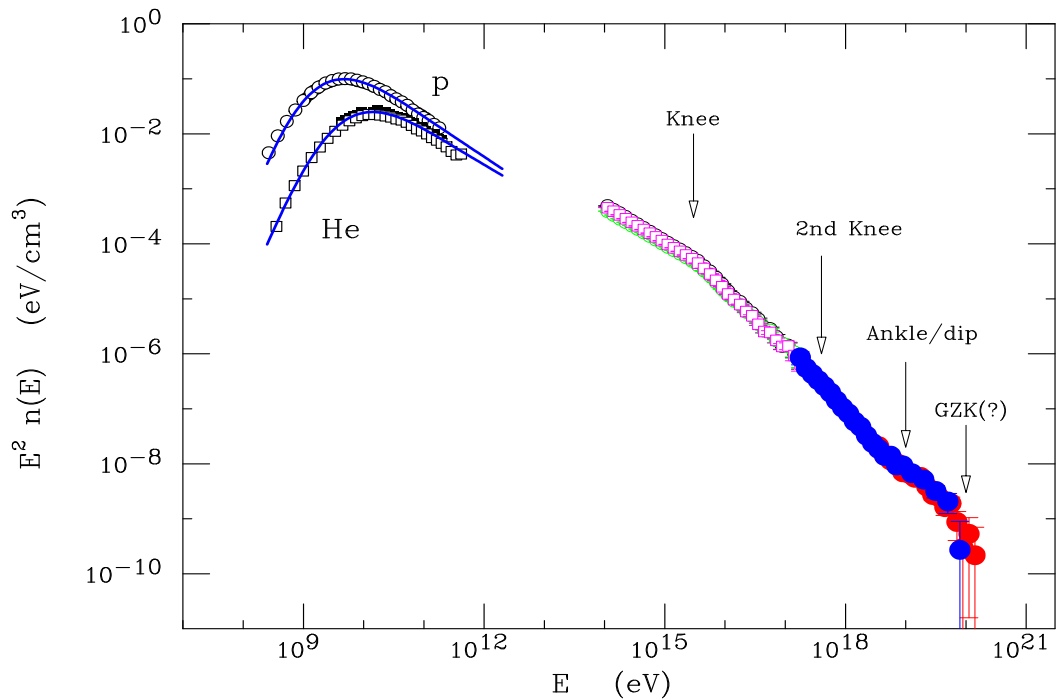,angle=90,width=14.cm,height=9.3cm}}
\caption {\footnotesize General structure of the CR energy spectrum
plotted in the form  $\phi(E) \, E^2$ versus  $\log E$
\label{fig:dens_cr} }
\end{figure}

 The spectrum of  hadronic CR  observed at the Earth
is  shown  in fig.~\ref{fig:dens_cr},   it  spans  approximately  
11   order of magnitude in energy
up  to $E \sim 10^{20}$~eV.   The CR spectrum  
is remarkably smooth and
can be reasonably 
  well  approximated by  a power law form ($\phi \propto E^{-\alpha}$) 
with a nearly constant slope.
The most prominent  spectral   features are the ``Knee'' 
at $E \simeq 3 \times 10^{15}$~eV, where the spectrum
steepens from a slope $\alpha \simeq 2.7 $ to   $\alpha \simeq 3$,
and  the ``Ankle'' at  $E \simeq 10^{19}$~eV, where 
the spectrum  flattens to  have again the  slope   $\alpha \simeq 2.7$.
At $E \sim  6 \times 10^{19}$~eV 
there is  now  clear evidence
(from the HiRes\cite{Abbasi:2007sv}
and Auger\cite{auger_merida_spectrum}  collaborations)
of a sharp  steepening of the spectrum. 
Such a  spectrum  suppression  had  been   predicted  
more that 40 years  by  Greisen, 
Zatsepin and Kuzmin\cite{GZK}  (GZK) as the consequence
of  the interactions  of high energy particles
with the photons of the 2.7~Kelvin Cosmic Microwave Background Radiation
(CMBR).  The non  observation of this   effect 
 generated  intense  interest and an abundant  literature.
The observed  spectrum  suppression
is  consistent  with the shape predicted 
for the GZK  mechanism  (pion production in  $p\gamma$ interactions),
 but other dynamical  explanations  are possible
including  fragmentation  of heavy nuclei, or an intrinsic
high energy cutoff in the  CR  sources. 
An  additional 
(less evident) steepening
of the  spectrum  (with the slope increasing  from  $\alpha \simeq 3$ to  
approximately 3.3) the so called  ``2nd--Knee'' 
at $E \sim 4 \times 10^{17}$~eV
has   recently emerged   as potentially  very significant. 
Explanations  for the origin of the ``Knee'', 
and the ``2nd--Knee''  and the ``Ankle'' are still disputed, and are considered
as of  central importance for an understanding of the  CR.

It is  now  known\cite{Sreekumar:1993hr}
  that most of the CR  that  we observe near the Earth  are  ``galactic'',
they are produced by sources inside  our Galaxy, and   are confined
by   galactic magnetic fields 
to form a  ``bubble'' 
(or ``halo'') around the visible disk of the Milky Way
with a  form and dimension that are still  the object some  dispute. 
Extragalactic  space should also  be filled by a  (much   more tenuous)
gas of CR,     produced by  the ensemble of all  sources in   the universe
during their  cosmological  history. 

 All galaxies
(including ours)  should be    sources of CR, as   the  particles
that ``leak out'' of  magnetic  confinement 
are  injected into  extragalactic space,
however  it is possible  (and predicted)
that  there are   additional  sources of extragalactic CR  that
are not normal  galaxies like our Milky Way, but  special
objects like for example Active  Galactic Nuclei (AGN) or  colliding galaxies.
The   extra--galactic  component  is expected to 
have  a harder energy spectrum, 
 and should emerge above the foreground  of galactic CR  at  
sufficinently  high energy.
The determination of a  ``transition energy''  that  marks the 
point  where the extragalactic  CR component  becomes  dominant
is  a very important  problem (see discussion in  
section~4.3).

\section{Galactic  Cosmic Rays}
Ultrarelativistic charged particles  produced inside our Galaxy
remain  trapped  by the galactic magnetic field,
that has a  typical  strength\footnote{
The magnetic  field receives  approximately equal
contributions   from  a  ``regular'' component 
(that in the galactic  plane has  field  lines that run parallel
to the spiral arms)
and an irregular component generated by 
 turbulent motions  in  the interstellar medium.
Our knowledge of the field has 
still significant  uncertainties and is    very important 
fields  of research.}    $B \sim 3~\mu$Gauss.
The    Larmor radius of a charged  particle in a magnetic  field is:
\begin{equation}
r_{\rm Larmor} = 
\frac{E}{Z \, e \, B} \simeq \frac{1.08}{Z}~  
\left (\frac{E}{10^{18} ~{\rm eV}} \right ) ~
\frac{{\mu {\rm Gauss}}}{B} 
~{\rm kpc}
\end{equation}
(one  parsec is  $3.084 \times 10^{18}$~cm, or 
approximately  3~light years).
Magnetic confinement becomes  impossible when 
the gyroradius is  comparable
with the linear dimensions  of  the Galaxy.
This corresponds to  energy:
\begin{equation}
E \aprge  Z \, e\, B\, R_{\rm Halo} \simeq
 2.7 \times 10^{19}~ Z~
\left (  \frac{B}{3 \; \mu{\rm Gauss}} \right )
~\left (  \frac{R_{\rm Halo}}{10 \; {\rm Kpc}} \right )
\end{equation}
Most CR  particles 
have   much lower energy   and a   gyroradius  that is  
much smaller than the galactic size.
The motion of these  particles
can   be  well approximated 
as  a diffusive process controlled  by the 
random  component of  the  galactic  field.   
The time needed   for a CR particles 
to diffuse  out of the galactic  halo
depends on its  rigidity (indicated  in the following 
for ultrarelativistic particles as $E/Z$).
In zero  order approximation 
(a spherical homogeneous halo) the confinement time of a CR 
can  be estimated   as:
\begin{equation}
\tau \left ( \frac{E}{Z} \right )  \simeq \frac{R_{\rm Halo}^2}{D(E/Z)}
\end{equation}
where  $R_{\rm Halo}$ is the  halo radius  and  
$D(E/Z)$  is a rigidity dependent diffusion coefficient.
An important consequence is that 
(assuming   stationarity)  the  population $N_A (E)$ of
CR  of type $A$  and energy  $E$ in the Galaxy is:
\begin{equation}
N_{A} (E) \simeq   Q_A(E) ~ \tau (E/Z)
\label{eq:gal_spectrum}
\end{equation}
where $Q_A (E) $ is  the injection rate 
and $\tau(E/Z)$ is  the (rigidity dependent) residence time.
Therefore the observed CR spectrum has not the  same energy distribution
of  the  particles near their sources, 
but is distorted and steepened    by confinement  effects:

The  determination of the shape  of the 
``source spectrum''  $Q(E)$  and of 
the  length   and rigidity dependence of the confinement time
are  obviously important problems.
A    powerful method is the  measurement
of the spectra   of   ``secondary nuclei''
(such as  Lithium, Beryllium and  Boron) in the CR.
These nuclei  are very rare 
in normal  matter, but they are  relatively abundant in 
the CR  flux, because they are produced in  the spallation
of parent  nuclei such as Carbon and Oxygen, a process
where  the    nuclear  fragments   maintain the same velocity
and therefore  the same  energy per  nucleon as the incident nucleus.
The energy spectrum of  secondary  nuclei  is
 expected   to be proportional to the square of the confinement time
(its own, and of the  parent  particle).
The measurements\cite{engelmann}   indicate that while  ``primary'' nuclei
have spectra  $\propto  E^{-2.7}$, secondary nuclei
have steeper spectra $\propto E^{-3.3}$. 
The conclusion is that the   confinement time
decreases   with  rigidity  as $ \tau \propto E^{-\delta} \simeq E^{-0.6}$,
and therefore the injection spectrum  of CR at their sources
has a slope $\alpha_{\rm inj} \simeq \alpha_{\rm obs} - \delta \simeq 2.1$.
The  absolute value  of the confinement time
can be  estimated  from the  relative  amounts of  secondary and 
primary nuclei in the CR flux, with additional  information
 obtained  studying the abundance of unstable  nuclear
isotopes\cite{Be10} with  appropriate  lifetime
(in particular  the nucleus   Beryllium--10  with 
a  half like of 1.5~Million years) and is of order of 10~million years for 
a  rigidity of a few GeV.

The  power law  rigidity  dependence  ($\tau \propto  E^{-\delta}$)
of the confinement time, 
and of the  diffusion coefficient
 ($D \propto  E^{\delta}$) is  not unexpected, because  for 
a  turbulent   magnetic  field  one 
predicts   a  power law  form  for the diffusion coefficient
with a slope  that is a function of the power spectrum of the field
irregularities.
For   the theoretically favoured 
Kolmogorov turbulence  one  expects $\tau(E) \propto E^{-1/3}$.
This   results  would imply a  steeper  injection spectrum 
$\propto  E^{-2.37}$.   The  discrepancy  between this  theoretical
prediction   and  data is is  still under discussion.

\subsection{The ``SuperNova Paradigm''  for galactic Cosmic Rays}
A  confinement  time  of  a few  million years is in fact very short
with respect to the age of the Galaxy, and therefore CR must continuously
be produced in our Galaxy (that is   in a stationary state
with approximately equal  CR injection and loss).
The  power of the ensemble of CR sources
in the Milky Way can  be estimated   as
the ratio  between the total  energy  
of   CR  in the Galaxy
divided  by  their average confinement time:
\begin{equation}
L_{\rm cr}^{\rm Milky~Way}  \simeq 
\frac{ {\rho_{\rm cr} (x_\odot)} ~ V_{\rm eff} } 
{\langle \tau_{\rm cr}  \rangle} 
\simeq 
 2 \times 10^{41} 
~\left ( \frac{\rm erg}{\rm s} \right )
\label{eq:galcrlum}
\end{equation}
For this numerical estimate  we have used   a  
local  CR  energy density\footnote{
The  estimate
of the CR  energy  density  is of the same  order of magnitude
of the  energy density of the  galactic magnetic  field
(for  $B \simeq 3~\mu$Gauss  one has $\rho_B = B^2/(8 \pi) 
\simeq 0.22$~eV/cm$^3$). This  is not a  coincidence, and  has a transparent
dynamical  explanation. The   galactic  magnetic  field
is  mostly generated by    electric currents    transported by the CR, 
while the CR themselves  are  confined by  $\vec{B}$, and therefore the 
two  quantities  are  close to  equilibrium.}
$\rho_{\rm cr} (\vec{x}_\odot) =  1.6$~eV/cm$^{3}$,
an effective volume  of 170~Kpc$^3$ and  a
average  confinement time of 20 Million years.
This   surprisingly  large    power  requirement
(approximately 50 million  solar luminosities),
is  a very important constraint for the identification
of   the   galactic  CR sources. 
In fact, simply on the basis of this  energy  condition
SuperNova (SN) explosions were proposed\cite{zwicky,Ginzburg:1969}
 as the  most likely CR source.  
The average  kinetic  energy of  a  SN  ejecta  is   of order
$L_{\rm kin} \simeq 1.6 \times 10^{51}$~erg 
(this corresponds  to the  kinetic  energy
of 10~solar masses  traveling at $v \simeq 4000$~Km/s) .
For  a SN rate in our  Galaxy  of order  every (30~years)$^{-1}$
the power  associated to  all SN  ejections  is  then
$L_{\rm SNR}^{\rm kin}  \simeq  1.7  \times 10^{42}$~erg/s, and
it is  therefore  possible to  regenerate the  Milky Way  CR if
approximately 10--20\% of the kinetic  energy of the ejecta
is  converted  into  a population of  relativistic  particles.

In  addition to these energy balance  considerations,
in the 1970's  a  dynamical  argument  emerged 
in favour of the SuperNova hypothesis,  when it was  understood
(see Drury\cite{drury}  for a review)
that  the  spherical  shock waves  produced  in the interstellar medium
by the  (supersonically moving) SN  ejecta   can  provide 
the environment where  particles, extracted  
from the tail of a  thermal  distribution,
are  accelerated  up to very high energy   generating
a power law  spectrum with a  well defined  slope
of order $\alpha \simeq 2 + \epsilon$  with   $\epsilon$ a small
positive number. That is  precisely  the  injection spectrum
needed  to  generate the  observed  CR.

The  basic  concept behind this  theory
is  an  extension of the   ideas  developed 
by Enrico  Fermi\cite{Fermi:1949ee}, who in 1949 made the  hypothesis
that  the acceleration of cosmic rays   is  a  stochastic process,
where each   CR particle 
acquires its  high energy in
undergoing many collision with   moving  plasma  clouds.
The clouds  carry (in their own rest frame)    turbulent magnetic fields
and  act as  `magnetic mirrors''  imparting on  average 
a positive ``kick''  to the scattering particle   with 
$\langle \Delta E \rangle /E \propto \beta^2 $
(with $\beta$ the cloud velocity).  It is easy to  see  that
this  process  generates a power  law   spectrum, 
with higher energy particles
having performed a  larger number of  collisions.
 
The crucial  new element  introduced in the 1970's
is the presence of the shock   wave.
Charged particles are  now  accelerated 
scattering against magnetic (in  the plasma  rest frame)
irregularities present  
both  upstream  and   downstream  of the  shock  front
that act as  Fermi's clouds, or  converging   magnetic  mirrors. 
The  new geometry   allows a   more efficient 
 acceleration
[$(\langle \Delta E\rangle/E)_{\rm crossing} \propto \beta_{\rm shock}$]
and  the  (mass, momentum and energy conservation) 
constraints  of  the fluid properties
across the  shock  determine the slope ($\alpha \simeq 2 + \epsilon$)
of the accelerated  particle spectrum. 

This  mechanism  (diffusive  acceleration in  the presence
of shock waves)  can  operate   every time that one has
a  shock  wave in an  astrophysical  fluid. The spherical
blast waves of SN  ejecta  are one  example  of this  situation
but several  other are known to exist.  In fact shocks
are  generated  every time  that  macroscopic  amounts
of   matter move  at supersonic  speed.
Particularly interesting  case are the jets  emitted
by Gamma Ray Bursts, 
by accreting   Black Holes    of stellar mass  (microQuasars)
or  supermassive (Active  Galactic  Nuclei). In all these objects there
in in  fact evidence  for  particle  acceleration.

The ``SuperNova Paradigm'' has a simple and  very important implication:
in the  vicinity of  young
SuperNova Remnant (SNR) one should find a population 
of relativistic  hadrons  (protons an nuclei)  with a spectrum
close to the injection one ($E^{-(2 + \epsilon)}$)   and a total energy
of order  $\sim 0.2 \times 10^{51}$~erg.
These relativistic particles    can  interact with the interstellar
medium  around the SN 
producing   neutral and charged pions 
that then  decay  generating photons  and neutrinos
with a spectrum  that to a good approximation
follows the same  power  law   of the parent proton  spectrum.
Describing the relativistic  proton  population
as  $N_p(E) \simeq  K_p~E^{-\alpha}$,
and approximating the confinement volume  as  homogeneous 
and with density $n_T$,  the  rate of emission of photons  and
neutrinos   is approximately:
\begin{equation}
\dot{N}_{\gamma(\nu)} (E) = K_p ~\left (c\,  \sigma_{pp} \, n_T \right )
~Z_{p \to  \gamma(\nu)}(\alpha)~E^{-\alpha}
\end{equation}
where $\sigma_{pp}$ is the $pp$ interaction cross 
section\footnote{For simplicity
we  use a  notation    that  indicates  only the dominant 
components  for  both the CR and the target medium.}
and
 $Z_{p \to  \gamma(\nu)}$ are adimensional  proportionality factors.
For a proton spectrum $\propto  E^{-2}$ that 
has  equal energy per  decade of energy the
$Z_{p \to \gamma(\nu)}$  factors are well  approximated
by the  (approximatgely energy independent)
fraction of the energy of the interacting protons
that  goes into  photons (or neutrinos), with 
  $Z_{p\gamma} \sim Z_{p\nu} \sim 0.15$.
Using the theoretically expected  slope $\alpha \simeq 2$, 
  fixing the constant $K_p$  by the  total    
energy contained in  the  relativistic
particles, assuming a  spectrum  extending 
in the  interval $[E_{\rm min}, E_{\rm max}]$,  a source
at a distance $d$ and integrating above
$E_{\rm th}$  the   photon (or neutrino)   integrated  flux at the
Earth is:
\begin{eqnarray}
\Phi_{\gamma(\nu)} (E_{\rm th})  & \simeq  &
\frac{\left (c \, \sigma_{pp} \, n_T \right )}
{4 \, \pi \, d^2} ~ \frac{E_{\rm cr}^{\rm tot}}
{\ln (E_{\rm max}/E_{\rm min})}
\, \frac{Z_{p \to \gamma (\nu)}} {E_{\rm th}} \nonumber \\
& ~ & ~ \nonumber  \\
& \simeq & 
0.9 \times 10^{-11} 
\; \left (\frac{\rm Kpc}{d} \right )^2 
\, \left (\frac{n_T}{{\rm cm}^{-3} } \right )
\, \left (\frac{E_{\rm cr}^{\rm tot}}{10^{50} \,
 {\rm erg} }\right ) 
\, \left (\frac{\rm TeV}{E_{\rm th}} \right )
\,
{\rm cm}^{-2}~{\rm s}^{-1}~~~
\label{eq:gnuflux}
\end{eqnarray}
whith  $d$ the source distance.  
Equation  (\ref{eq:gnuflux}) is a  key prediction of the 
``SNR  paradigm'' for the acceleration
of the bulk of cosmic  rays that  can  be tested
with  photon  observations  in the  GeV/TeV  regions,  and
soon  also  with  neutrino  telescopes.

In  2004\cite{Aharonian:2004vr}
 the HESS   TeV  Cherenkov  telescope observed the SuperNova Remnant
RX~J1713.7-3946\footnote{
An object discovered in  X--rays  by the ROSAT  instrument and 
then associated,  with a  very high  degree of  confidence, to
a supernova   observed and  recorded
by  chinese astronomers in  the  year  393 of the current era.}
as a  very bright  source of  TeV photons.
The property of the   photons 
from this source, taking into  account
our knowledge of the density
of the environment around the object
 are  consistent\cite{Aharonian:2006ws} with
the  expectation of the ``SNR paradigm''   for  galactic cosmic  rays.

Several other young SNR  have also been detected
with  TeV photons  (this includes   the detection  of 
Cassiopea~A\cite{Albert:2007wz}),  
while for other objects one  has only upper limits.
These  observation of SNR   in high energy photons
gives  support to the ``SNR  paradigm''  for CR acceleration  
(for a more critical view
of the situation see\cite{DeRujula:2006qq}). 
This conclusion  is however not  unambiguous; also for the 
best candidate source RX~J1713.7-3946
a  ``leptonic origin'' of the radiation
(inverse Compton  of  relativistic  $e^\mp$  on the  radiation fields
around the SN)  cannot be entirely excluded.

It is  remarkable that the neutrino  flux   that should 
accompany the photon flux  in case of a  ``hadronic''  origin
($\pi^\circ \to \gamma \gamma$  decay, accompanied by 
$\pi^+ \to \mu^+ \, \nu_\mu \to (e^+ \, \nu_e \, \nubar_\mu) \; \nu+\mu$ 
chain decay)  is  detectable,  at  the level of  few 
neutrino  induced  up--going muon events per year, in a 
Km$^3$ size  neutrino telescope  placed in the northern  hemisphere.
Photon  observations  in the GeV region with detectors
on satellites  (Agile and GLAST)  have  also the potential
to  test the  ``SuperNova paradigm''  for CR production.

\subsection{The ``Knee'', and  the maximum acceleration energy}
The mechanism  of  diffusive particle acceleration  generates
a power law spectrum only up to a finite  maximum energy $E_{\rm max}$.
This  maximum  energy is  determined  by the  product
of the   acceleration rate 
times the total  time  available for the acceleration.
Simple   considerations allow to estimate
 $E_{\rm max}  \simeq  Z \, e \, B \, R \;  \beta_{\rm shock}$
where $B$ is the typical strength of the magnetic  field, $R$ 
the linear dimension of the acceleration site  and $\beta_{\rm shock}$ the
shock velocity.  Note  that this estimate is  more
stringent  than a simple  magnetic containement  condition
by a  factor $\beta_{\rm shock} < 1$. 
Substituting values  that are relevant for  a  SNR  one finds:
\begin{equation}
 E_{\rm max}  \simeq Z \, e \, B \, R \;  \beta_{\rm shock}
\simeq
0.4 \times 10^{14}~{\rm eV} ~Z~
\left (
\frac{R}{5~{\rm pc}}
\right )
~
\left (
\frac{B}{3~\mu{\rm Gauss}}
\right )
~
\left (
\frac{\beta_{\rm shock}}{0.03}
\right )
\label{eq:emax}
\end{equation}
This  estimated  maximum  energy is  much smaller
than  the highest observed  energies,  and therefore
it is  clear than   other  type(s) of CR  sources  must exist.

The estimate 
(\ref{eq:emax}) is however not  too  dissimilar
from the   ``Knee'' energy  (at $E \simeq  3 \times 10^{15}$~eV, 
and it has  been suggested  that the  ``Knee structure'' 
marks in fact the maximum  energy for  {\em proton}  acceleration in SNR.
The apparent  smooth  behaviour of the spectrum above  the knee
could hide  a succession of cutoffs 
with increasing energies
($E_{\rm max}^{Z} = Z \; E_{\rm max}^{p}$)
for heavier  nuclear species.
In this  model the SNR spectrum would extend  up to the 
maximum energy for  iron  
with  $E_{\rm max}^{\rm iron} \simeq 0.8 \times 10^{17}$~eV.
The key  prediction here is the gradual   change 
(with increasing  $\langle A \rangle$)
CR compositon  above the knee, or more precisely a  set of different
``knees''  at identical  rigidity,  and therefore scaling in energy
proportionally to the electric charge $Z$.
Unfortunately  the measurement
of particle mass in Extensive Air Showers in the region of the knee\footnote{
The best method  is  the simultanous measurement of the 
electromagnetic  and muon component of the shower  at ground level.}
are not easy and  suffer from significant uncertainties in
the modeling of the shower (see below in section~6).
The data of Kascade\cite{kascade}
 and other experiment are (with large errors)  roughly 
consistent with this  idea.

Equation  (\ref{eq:emax})  for the maximum energy 
is valid in  general,  for all astrophysical environments where
shocks are present in  a diffusive medium,
and setting $\beta_{\rm  shock} = 1$ the equation 
 simply gives a  very general condition for  containement in the  source.
Since one observes  particles  with  energy up to $10^{20}$~eV,
the acceleration of these particles   requires  
sources   with  a sufficiently large  $B\; R$ product,
that is  sources   with  a sufficiently  strong magnetic  field
and sufficiently    extended  to  satisfy    condition
(\ref{eq:emax}). 
This  point has  been  stressed  by Michael Hillas\cite{Hillas:1985is}
that  initiated a  systematic  study of all possible
astrophysical environments  that  are possible
candidates for ultra high energy acceleration on  the basis
of  equation (\ref{eq:emax}).  The bottom line of this  analysis
is  that this  very simple  constraint  is  very ``selective''
and only  few  objects  are  viable for acceleration
up to $E \sim 10^{20}$~eV.   The main  candidates are
Active Galactic  Nuclei, and 
the jet--like emission associated with the Gamma Ray Bursts.

\subsection{Alternative Models}
A possible  alternative  to the  SNR paradigm  is 
the ``cannon ball model'' of Dar and De Rujula\cite{Dar:2006dy}.
In this model  the  Energy source  for the production of
CR is  ultimately the same as in  the  SNR paradigm, and is
the  gravitational  binding energy  released   during
the collapse of massive stars  at the end  of their
evolution,  but the  dynamics  of CR acceleration is deeply  different.
In the cannon ball model    the  explosion due to a gravitational
collapse  produce  in most of perhaps all of the cases
not only  a   quasi--spherical
mass ejection  with an initial    (non--relativistic)  velocity  of order
10$^{4}$~Km/s, but also  the emission along the  rotation  axis
of the   collapsing star  of  ``blobs'' of   ordinary matter 
each  having a mass of order $10^{26}$~grams  and
 an extraordinary  Lorentz  factor  of  order $\Gamma = 1000$;
the ``cannon balls''  emission, according to the authors is  responsible
for most  of the Gamma Ray  Bursts\cite{dar-derujula-grb}, 
with a mechanism  that differs  from the often discussed 
``Fireball  model''\cite{Piran-grb,zhang-meszaros}.
Cosmic rays  are accelerated
by the collisions of the cannon balls while they
decelerates slowly  colliding with the interstellar  medium.

In this  work  we do not have  the space to  discuss critically this model,
see  reference\cite{Hillas:2006ms}  for  criticism
 (and the authors' response\cite{Dar:2006bp}).

\section{Extragalactic Cosmic Rays}
Most ultrarelativistic  particles  injected in   extragalactic space
have  negligible    interaction  probability, and  only lose energy
because of the universal  redshift effects.  The important 
exceptions  are those  particles that   have a
 sufficiently high energy to interact with the 
abundant photons  of the Cosmic   Microwave Backround Radiation
(CMBR).
For protons  one  has  to consider two  important    thresholds
associated  to  pair production      ($p  + \gamma \to p + e^+ e^-$):
\begin{equation}
E_{\rm th}^{e^+e^-}  \sim  \frac {2 \, m_e \, m_p}{10 \;  T_\gamma}  \simeq  
4 \times 10^{17}~{\rm eV}
\label{eq:eth1}
\end{equation}
and  pion  production ($p  + \gamma \to N+ \pi$):
\begin{equation}
E_{\rm th}^{\pi}  \sim  \frac { m_\pi \;  m_p}{10 \;  T_\gamma}  \simeq  
6 \times 10^{19}~{\rm eV}
\label{eq:eth2}
\end{equation}
For  heavy nuclei one has  to consider
the  photodisintegration     reactions    with threshold
 (related to the process $A + \gamma \to  (A-1) + N$):
\begin{equation}
E_{\rm th}^{\gamma A} 
\sim  \frac { A\,  m_p \; \varepsilon_B}{10 \;  T_\gamma}  \simeq  
3 \times 10^{18}~A ~\left ( \frac{\varepsilon_B}{8~{\rm MeV}} \right )
~{\rm eV}
\label{eq:eth3}
\end{equation}
where $\varepsilon_B$ is the binding  energy of the
displaced nucleon. Note that the threshold  energies
for protons  and iron   nuclei ($A\simeq 56$)  differ
by a  factor 2--3. 
Particles  above  the thresholds  (\ref{eq:eth1})
and (\ref{eq:eth2})   lose rapidly  energy
cannot propagate  only for a  cosmologically short time.
This  results  in a strong  flux  suppression.

The  (space)  average extragalactic  density  $\langle n_a(E) \rangle $
of particle type  $a$  (for example  protons, or neutrinos) 
can be calculated integrating  the source emission 
over the   entire  history of the universe,
taking also into account the effects of  energy loss or absorption.  
A  general  expression  for the average particle density is:
\begin{equation}
\langle n (E) \rangle  = \int_0^\infty dt~ \int dE_g
~  q(E_g, t)   ~ P(E,E_g,t)
\label{eq:phi_calculation}
\end{equation}
where    the double integral is over the emission  time and the emission
energy,  $q(E,t)$ is the (space  averaged) emissivity 
for  the particle  considered (that is 
the number of particles emitted  per unit  of time, comoving  volume
and energy at universal  time  $t$),  and
$P(E,E_g,t)$ is the  probability  that a particle
generated  with energy $E_g$    at time $t$  
survives  until the present  epoch with energy $E$.

Expression (\ref{eq:phi_calculation})  can be  simplified
if one  makes the assumption that the energy loss is a 
continuous  deterministic  process, and  the 
probability $P$ can be approximated with the expression:
\begin{equation}
P(E,E_g,t) = \delta [E -E_g(E,t)]
\end{equation}
where the function $E_g(E,t)$  is the energy of a particle
observed now with energy $E$ 
 after propagation backward in time to
time $t$.
This approximation is essentially perfect
for neutrinos, and also sufficiently accurate for protons.

One can rewrite equation
(\ref{eq:phi_calculation})  changing  the integration
 variable from $t$  to  the  redshift $z$ 
with the Jacobian factor:
\begin{equation}
\frac{dt}{dz} = -\frac{1}{H(z) \, (1+z)}
\end{equation}
where  $H(z)$ the Hubble  constant at the epoch redshift $z$:
\begin{equation}
 H(z) = H_0 ~ \sqrt{\Omega_{\rm m} \, (1+z)^3
+ \Omega_{\Lambda} 
+ (1 - \Omega_\Lambda - \Omega_{\rm m}) \, (1+z)^2} ~,
\end{equation}
and  performing  the   integration over 
$E_g$  using the delta function.  
The result can  be  written in  the form:
\begin{equation}
\langle n(E) \rangle  =
\frac{1}{H_0} ~
 q(E,0) ~\zeta (E) 
\label{eq:shape_factor}
\end{equation}
where  $\zeta(E)$ is an adimensional ``shape  factor'':
\begin{equation}
\zeta(E) = \int_0^\infty dz~ \frac{H_0}{ H(z)} \; \frac{1}{(1 +z)}
~ \frac{  q[E_g(E,z)] }{ q(E,0)} 
~ \frac{ dE_g(E,z)}{dE} 
\label{eq:shape_factor1}
\end{equation}
The expression  (\ref{eq:shape_factor}) 
has a  trasparent physical   interpretation:
$q(E,0)$  tells  us  how  many particles
of energy $E$ are injected  per unit time   and unit  volume
in the present universe, multiplying by the Hubble  time
$H_0^{-1}$   we obtain  a  first order  estimate of the 
present density, and the ``shape  factor''  contains
all  complications  related  to
the  expansion of the universe,  the cosmological 
evolution of the sources,  and the   energy
losses  of the  particle  considered.

\subsection{The proton  extragalactic spectrum}
We can now  use expression  (\ref{eq:shape_factor}) 
to compute the extragalatic  proton flux
assuming that it is  related to the space
averaged  density by the simple  relation\footnote{In general
the effects of the magnetic  field, and of the ``granularity of
the sources  can result in modification of this  assumption.}
\begin{equation}
\phi_p (E) = \frac{c}{4 \, \pi} ~ \langle n_p(E) \rangle
=   \frac{c}{4 \, \pi} ~ \frac{1}{H_0} ~q(E,0) ~\zeta(E)
\label{eq:phi_dens}
\end{equation}
The simplest assumption for the  injection is  a  
power law   form:
$q(E)  = q_0 \; E^{-\alpha}$  that depends only on  two parameters, the
slope $\alpha$ and the  absolute normalization.
The calculation  of the flux has three  obvious steps,  the first 
is the calculation  of the  evolution of the proton energy with redshift;
the second   is the calculation of the  shape  factor $\zeta(E)$
according  to  equation (\ref{eq:shape_factor1}), and 
finally one can  use   equation  (\ref{eq:phi_dens}).

\begin{figure} [hbt]
\centerline{\epsfig{figure=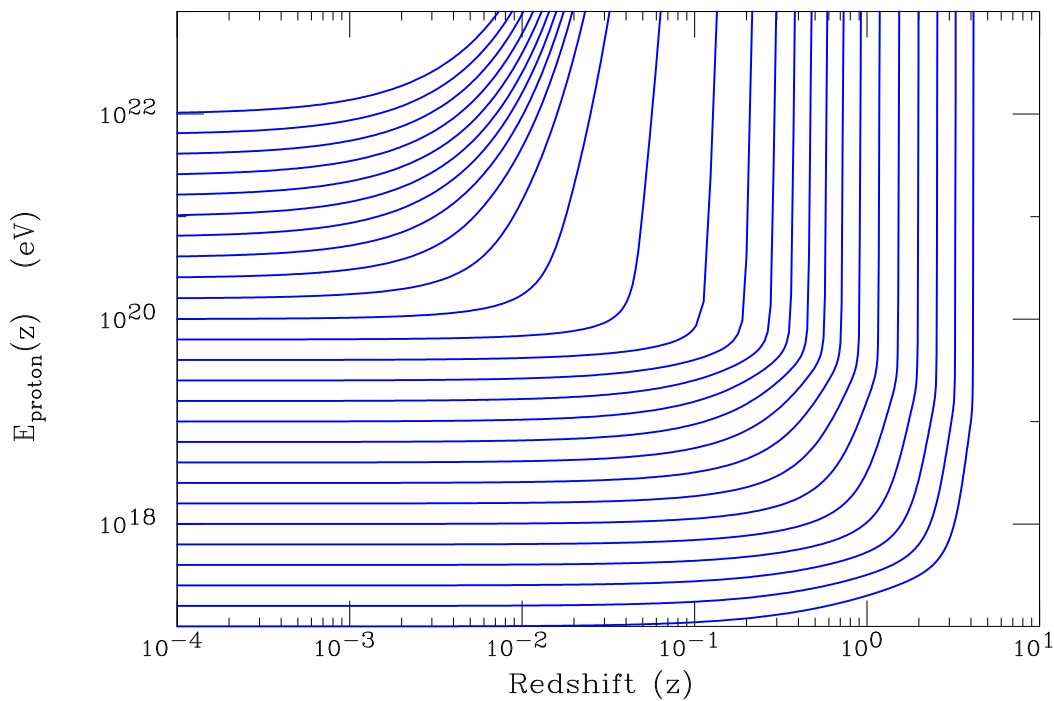,angle=90,width=14cm,height=9.3cm}}
\caption {\footnotesize 
Redshift (time)  dependence of the  energy of protons
observed   with     different  final energy at the present epoch.
\label{fig:e_evol}  }
\end{figure}

The results  of a  numerical calculation  
of $E_g (E,z)$ are  shown in fig.~\ref{fig:e_evol}.
The different  curves   correspond to the 
``backward in time'' evolution  of the energy of protons  detected
``now'' with $E$
between $10^{17}$~eV  and $10^{22}$~eV.
For  each  energy $E$ one  finds  (at a certain redshift $z$)
a  sort  of  ``photon wall'',
where the $E(z)$  starts to grow    faster  than an  exponential. Clearly
at this  critical redshift the  universe
(filled with a hotter and denser radiation field)
has  become  opaque  to the propagation  of the  protons.
Note that the   time evolution of the energy of 
a  proton   is   completely independent  from the 
structure and  intensity of the  magnetic  fields,  because  
a magnetic  field can  only  bend the trajectory of a  particle 
and  the   target  radiation field 
is  isotropic\footnote{Clearly in this case the redshift mantains
its one to one correspondence with the time of emission of the particle,
but loses its  simple  relation with the space  distance
of the particle source.}. 

\begin{figure} [hbt]
\centerline{\epsfig{figure=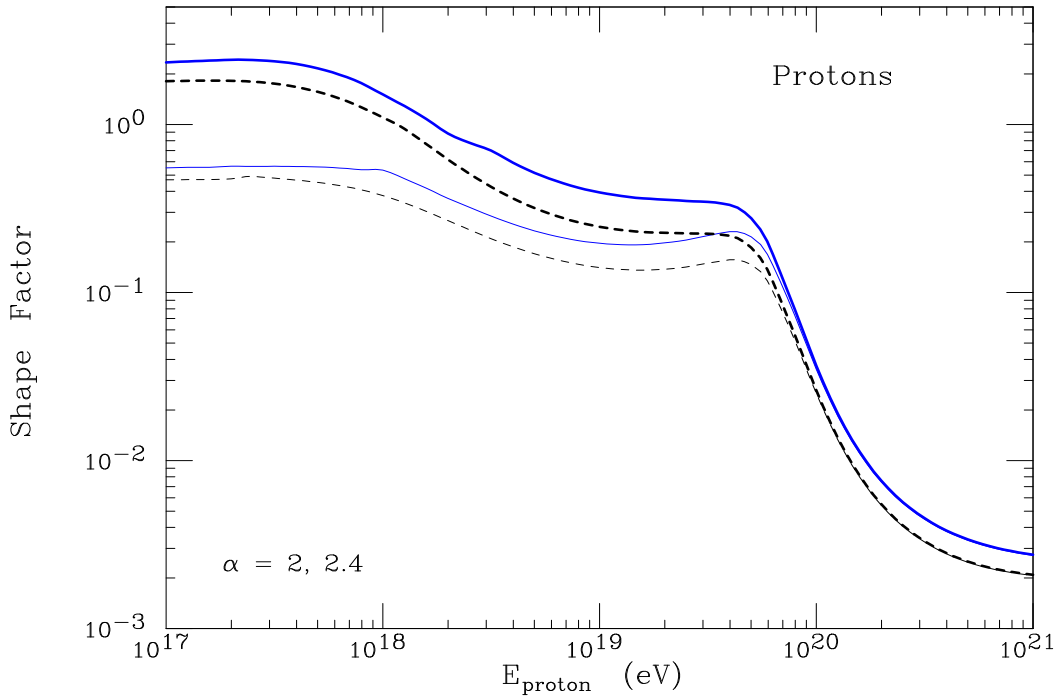,angle=90,width=14cm,height=9.3cm}}
\caption {\footnotesize 
Shape  factor  $\zeta(E)$  for  protons, calculated  for
a  power  injection with  slope
$\alpha = 2.0$, 2.2, 2.4 and~2.6.
 The solid lines are calculated  for a constant value
of the  source  power density, while the dashed  lines assume
for ${\cal L}_p(z)$ the   redshift  dependence  shown 
estimated by  Ueda  \protect\cite{ueda} for the X--ray  luminosity of AGNs.
\label{fig:csi_p}  }
\end{figure}

The result of the calculation of the ``shape factor''  $\zeta(E)$ for
a power law injection is   shown  in fig.~\ref{fig:csi_p}. 
It is  important to  note  that the   
function  $\zeta(E)$   has   a non trivial energy dependence,
and   while the injection spectrum is  a smooth power law  will show
some  structure,  the  ``imprints'' of the energy 
loss  processes (pair and pion production)   of the protons.
For energy  $E \ll  E_{\rm th}^{e^+e^-}$, only    adiabatic 
(redshift)  energy losses are  significant,
and in this case the  
the shape  factor $\zeta(E)$ tends to a constant  value:
\begin{equation}
\zeta   \to  \int_0^\infty dz~ \frac{H_0}{ H(z)}
(1+z)^{-\alpha} ~
~ \frac{{\cal L}(z)}{{\cal L}_0} 
\label{eq:shape_nu}
\end{equation}
(${\cal L}(z)/{\cal L}_0$ is the 
cosmological evolution of the source luminosity).
Accordingly, the  power  law  injection 
results  in an observable  flux that is  a power law
of the same slope.
At higher  energy  the energy losses due to  the pair production  
and pion production   thresholds
(equations (\ref{eq:eth1}) and (\ref{eq:eth2}))
leave their ``marks'' on the shape of the  energy spectrum.

\begin{figure} [hbt]
\centerline{\epsfig{figure=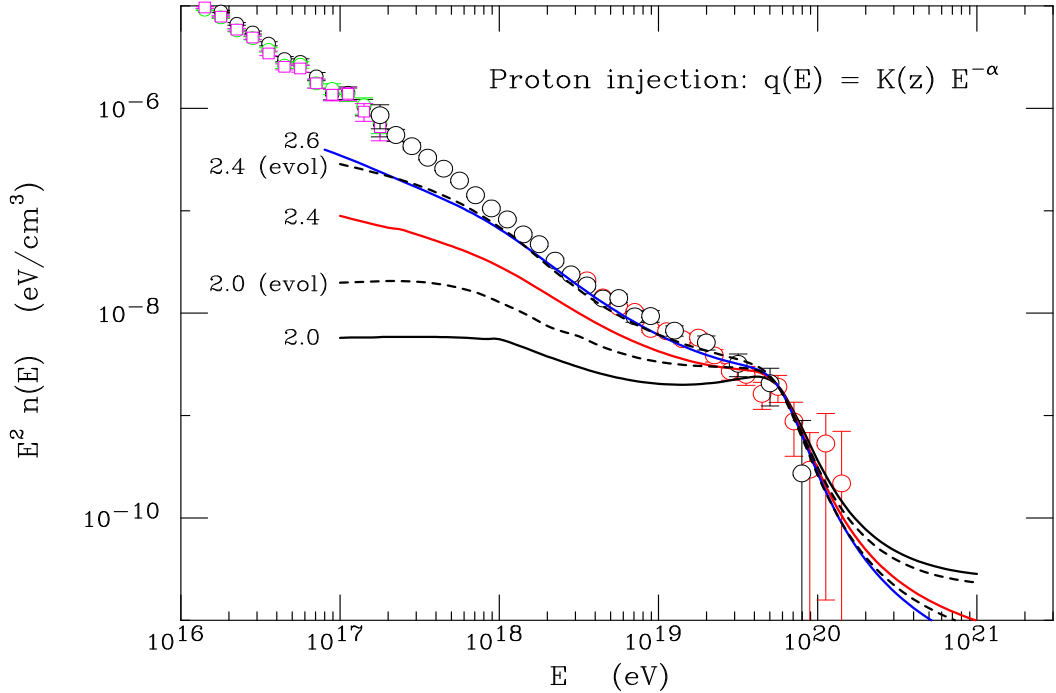,angle=90,width=14cm,height=9.3cm}}
\caption {\footnotesize   Extragalactic  proton  spectrum
calculated for a power law injection 
with different  slopes  $\alpha$, for  no--evolution of 
and with the evolution of\protect\cite{ueda}.
The  data points are from  the  Tibet array\protect\cite{tibet}
and the   HiRes detector \protect\cite{Abbasi:2007sv}.
\label{fig:densa}  }
\end{figure}

\begin{figure} [hbt]
\centerline{\epsfig{figure=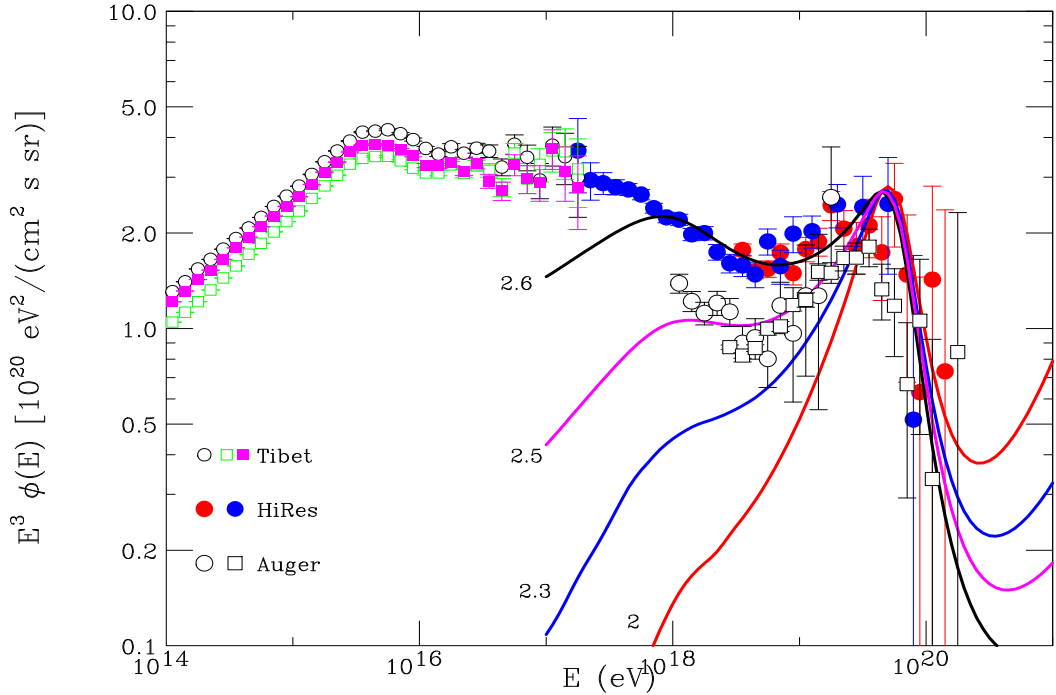,angle=90,width=14cm,height=9.3cm}}
\caption {\footnotesize   High energy CR spectrum.
For clarity only the data of the Tibet array\protect\cite{tibet}
  (below  $E\simeq 10^{17}$~eV),
  HiRes\protect\cite{Abbasi:2007sv}
and Auger\protect\cite{auger_merida_spectrum}
 (at higher energy) are shown.
The spectrum is   plotted  
in the form $\phi(E) \, E^3$ versus  $\log E$.
On can  note  the different  energy scales for
the Hires and Auger experiment.
The Tibet data points  are calculated using three different
interaction models for   the reconstruction of the shower  energy.
The lines  describe
the shape of an extragalactic  components  with injection spectrum
$E^{-\alpha}$   and  normalized to  fit   the high energy data.
\label{fig:uhecr}  }
\end{figure}

The results  of the calculations for the  obervable 
flux are  shown in  fig.~\ref{fig:densa} and fig.~\ref{fig:uhecr}.
The  calculations were   performed
with  different slopes of the injection power law
$\alpha = 2.0$, 2.2, 2.4 and 2.6, and with two different
assumptions for the evolution of the cosmic  sources,
a  constant injection, and  evolution  equal
to the one   estimated in\cite{ueda}  for the AGN  hard X--ray
emission. The  normalization of the  calculation is chosen
in order to have a good fit to the observed data in the 
very high energy  region.
 Some   comments about the results  shown in fig.~\ref{fig:densa}
are in order:

(i) The   shape of the  observed  flux   above
$E \aprge 5 \times 10^{19}$~eV  is 
consistent with  the presence of 
the suppression predicted  by GZK\cite{GZK}.
The implication of this point are discussed  below in section~4.2.

(ii)  The shape  of the calculated   spectrum   is determined
by  the  slope $\alpha$  of  the injection spectrum and
also by the  assumptions  about the  redshift dependence of the injection.
Actually this  redshift dependence
is  of small importance  for  $E \aprge 5 \times 10^{19}$~eV  
 because  the lifetime of a very high energy particle is    brief
and  they do  not probe  deeply into the  past history of the universe.
For  lower energy the   inclusion of source evolution 
is  significant. Lower energy particles
can arrive  from a more distant past  when   the  emissivity
was  higher, and  therefore, for the same 
shape of the injection,  the inclusion of
source evolution  results in a softer  spectrum.
As a rule of  thumb, including  for the injection
the redshift dependence of   the AGN  luminosity
(or also of the  stellar formation rate)
has an effect similar to  the softening of the
injection spectrum of order  $\Delta \alpha \simeq 0.2$. 

(iii)  It is  remarkable  that if one  choses  an injection spectrum
with a shape $ \propto E^{-2.6}$   and  no source evolution (or a source
$ \propto E^{-2.4}$   with  cosmic  evolution)  with   appropriate
normalization, the calculated  flux  can provide  a good
match to the data  in the entire region from
$E \simeq 10^{18}$~eV  up to the highest energies.
The possible implications of this
result are  discussed  below in section~4.3.

\subsection{The ``GZK  controversy''}
For  several years the  ``dominant   question''  in CR  physics  has
been the possible existence of  a  large flux  of particles
with energy   larger that  $10^{20}$~eV. 
To  solve this  very surprising  result  two main 
arguments   have been proposed.
One possibility is the  existence of an additional  source
of particles of  extremely high  energy    from the decay
of super--heavy 
(mass of order of the GUT  mass $M_{\rm GUT} \simeq 10^{25}$~eV).
A second  possibility   is the existence of violation  of the
 Lorentz  invariance. The process  $p\gamma \to N\pi$ that is at the
basis of the expected suppression of the proton flux  is in fact
studied in the laboratory  mostly in the $p$ rest frame, while
for  CR it is  relevant in  a frame where the $p$  has  a Lorentz 
factor  $E_p/m_p \sim 10^{11}$. One can therefore 
speculate that ``small''  violations
of the Lorentz invariance could  suppress the cross section for  high 
energy CR.
Ultra High Energy  Cosmic  Rays therefore offered the 
exciting possibility  of being the key observational field
for  the exploration for new  physics.

The new  results of the HiRes\cite{Abbasi:2007sv}
and Auger\cite{auger_merida_spectrum} collaborations
demonstrating the existence
of a suppression of the CR flux at 
$E \simeq 6 \times 10^{19}$~eV
have however    very significantly changed  the ``scientific  landscape''
around  the field. The phenomenological motivations for ``exotic physics''
in  UHECR   have essentially disappeared. 
Of course, the theoretical
arguments  behind the existence
of  supermassive topological defects    as  relics from  the early 
universe,  and   (perhaps more controversially) of possible violations of  
the Lorentz invariance remain in existence, and therefore 
UHECR can be used to set limits, or look for hints
of these phenomena.  However the field of UHECR   has now  returned
to  its status  as an  ``astrophysical  problem'',  where the main
questions  is to establish what are the   sources of these 
ultrarelativistic particles,  
and what are the physical processes responsible for  the acceleration\footnote{
{\it A posteriori} it could  be an interesting problem
to  consider  why essentially the entire  community of CR physicists 
has remained  so ``captured''   with the  idea  of the existence of
``super--GZK''  particles''  on the  basis of an evidence  of
weak statistical significance  and with the possibility
of significant systematic errors. The ``temptation'' of the possible
discovery  of a result of  extraordinary importance   probably played 
a  significant role  for  both  observers and theorists.}.

Does  this  new situation makes the study of CR less  interesting?
In a certain sense the answer is  obviously yes.  The  significance
of the ``exotic physics''  for the UHECR   would clearly be of
truly revolutionary significance.  On the other hand the astrophysics
of CR, or  more in general
study of the origin of the different 
forms of high energy radiation  remain  a field of
{\em  fundamental} importance.

The  development of  fundamental  Science has   historically
always been intimately  connected with the discovery  and
understanding of astrophysical  objects. Steps in this remarkable
history have been the   understanding of: the structure  and  dynamics 
of the solar system; the source of energy
in the Sun and the stars;  the quantum mechanical
effects that  sustain White Dwarf stars;  the formation of neutron stars
and their connection with SuperNova explosions.

The sources of the high energy radiation in the universe 
(some of which  are  still  undiscovered)  are    
very likely (together  with the early universe) the most
``extreme''  environments   where we can perform the most
stringent tests   for  physical  laws in conditions that  are often
simply not achieavable in Earth based laboratories.
In particular, the regions  of space near compact objects or near the horizon
of Black Holes are very likely particle  acceleration sites.   
It is  also possible  (or perhaps  likely)   that the still 
unexplained  cosmological  Dark Matter  will play an important  dynamical  role
in precisely these acceleration sites 
(for  example  near the center  of galaxies), and  therefore
one  could  have to   disentangle  the physics of particle acceleration
and Dark Matter annihilation. 
In a nutshell:  High Energy Astrophysics    remains a
crucial  field for  fundamental Science.

\subsection{Galactic to Extragalactic  transition.}
 Berezinsky and  collaborators\cite{Berezinsky:2005cq}
have been the  first to note the
remarkable  fact  that the injection of a smooth   (power
law) spectrum of protons  (with a  sufficienly small
contributions of helium and other nuclei)  can  describe  the UHCR 
not only above the ``ankle'' but 
in the entire energy range  $E \aprge 10^{18}$~eV.
These authors  stress  the point that
the  prediction   depends on  only   few parameters
(the slope $\alpha$, the injection absolute normalization
 and  the $z$ dependence of its evolution)  can provide a  very good fit
to the data,  and have concluded  that this  
is very unlikely to be a simple   coincidence,
and   is evidence of  a  real  effect. 

The  ``Ankle''  feature in the CR spectrum
has  been   for a long time
assumed to mark the transition  energy     where 
the softer  galactic  component   is overtaken  by 
the  harder  extragalactic one.
In  the Berezinsky et al. model the  flattening
of the  spectrum    emerges from an entirely different   physical  mechanism.
In the entire  energy range, both  above and below the ``Ankle'',
the spectrum is due to a single,  proton  dominated component,
injected with a   smooth (power law) spectrum.
The visible structure in the spectrum  is interpreted as the  imprint
of energy losses  due to   $e^+e^-$ pairs production. 
This effects is  mostly effective
at an   energy  range around $E \sim 10^{19}$~eV, and
form  a ``dip''  in the spectrum. The energy range of the effect
is controled  by the   kinematical  threshold for 
pair production at low  energy, and by the fact that
the energy lost for the production of an $e^+e^-$ pair
is  $\propto m_e^2/E$  decreases with $E$.
The shape  of the  sprectrum  distortion or  ``Dip'' 
is in fact determined  by  simple, fundamental physics   cosiderations 
and robustly predicted. The form of the spectrum is  of course
strongly  dependent on the value of the slope $\alpha$.
When  the slope is $\alpha \simeq 2.6$ 
(for no--evolution of the injection) 
or $\alpha \simeq 2.4$ (for  AGN/SFR source evolution)
the spectral shape  can  describe the data  in the
entire   energy range: $[10^{18}, 10^{20}]$~eV,

The name ``Dip model''  encodes  this new dynamical  explanation
and  renames the structure traditionally called  ``Ankle''.
In this  model  the  transition between  galactic  and extra--galactic
cosmic  rays  happens at  lower energy  and corresponds to  the
less prominent spectral  steepening  feature   called
``2nd--Knee''.  For more discussion about  observations
of this  structure see\cite{Bergman:2007kn}.

The ``Dip Model''   has  attracted  considerable  interest, and
is  a leading contender  for the  description of the
highest energy  cosmic rays.
In my view,  the  questions  of  where is
the  transition between  galactic  and extra--galactic cosmic rays,
and if the ``Ankle'' is really a ``Dip'' or viceversa  remain open,
at the center of an important   controversy, that will
be  soon   settled by improved  observations.
The implications  of this problem   are   important and
wide. 

If the ``Dip Model'' is correct,  life for the observers
would be significantly  simplified. The reason 
is that it allows  ``calibrations''   for the  measurements 
of both  energy  and  mass of the CR   starting from
$E \sim 10^{18}$~eV.
It is  clearly  possible to use an  observed  feature of the spectrum
(with a  well understood  physical origin) to 
determine the scale  of the energy 
measurements\footnote{Matching  the spectral shapes
of different experiments one can estimate
the  {\em ratios}  between their respective  energy scales.
To  establish the absolute energy scale
one  obviously need  to compare  the observed
spectra with a  ``template''  based  on 
an understood  physical mechanism.}.
In fact, the idea to use  the 
``GZK suppression''  to   fix the CR energy scale  is now  40 years  old.
In the ``DIP  model''   one  has a  spectral  feature  (the ``Dip'')
that allows to play the same game
at lower energy and with much higher  statistics. 
Berezinki  and  collaborators have already suggested 
corrections  factors for the energy scale of the different
UHECR experiment.
It should be streessed  that the validity of the ``Dip Model''  
depends on   the  fact that the energy scale of the HiReS  experiment
is approximately correct, while the energy scale of the Auger experiment
is  significantly  underestimated, as shown in fig.~\ref{fig:uhecr}.

The  same  discussion can be made 
for  the CR composition measurements.
An established ``GZK'' feature  implies  a proton  rich spectrum.
In the ``Dip Model''   most particles  above $E \sim 10^{18}$~eV are 
protons.  With this  knowledge one  can use the observations
on the   shower  development to  extract information 
on hadronic  interactions at high energy (see 
discussion in section~7).

The relatively ``soft'' injection
($\alpha = 2.4$--2.6, 
depending  on  the assumptions  about the
source resdshift  evolution) of  the ``Dip model'' is  not expected
in several acceleration models for both AGN's and GRB's
that predict   a  slope closer to $\alpha =2$.
It  should be  noted  however  that the
injection  considered  here is the space averaged  one,  
and  it is possible\cite{Berezinsky:2005cq}
that  the spectra of  {\em individual} sources
are  all  flatter (for example  with $\alpha \simeq 2$  but have
different   high energy cutoffs,  that combine 
to an average spectrum that can be  approximated 
with an  effective softer  power law.
This  explanation  is however  
problematic,  and  it is   unclear 
if it can  naturally result in 
spectrum with the required smoothness.

The determination of the galactic/extra--galactic transition energy 
has  obviously  important consequences  also  for the  properties
of the galactic   spectrum, and the galactic sources. 
In the models that  have the transition 
at the ``Ankle'' it is necessary to  assume that our own  Galaxy  contains
sources  capable to  accelerate
particles  up to $10^{19}$~eV and above.  In the ``Dip model''
with a lower energy transition
one has a  less  stringent  (by a factor 10 of more)
requirement for the maximum  energy of the Milky Way accelerators.

\vspace {0.2 cm}
The  transition energy between  galactic and 
extra--galactic  cosmic rays 
obviously marks the passage  from a softer to  a harder component.
It is therefore {\em natural} to expect that the  transition
appears as an {\em hardening}  of the spectrum.
The `Ankle'' is the only   observed
hardening of the spectrum,  in the   entire  CR energy range,
this has  lead   to  its  identification with the transition energy.
In  the Dip model  the 
 galactic to  extra--galactic  transition corresponds 
however to  the ``2nd--Knee'', that is to
a spectral {\em steepening}.
This  may appear, and indeed it is  surprising.
The    situation where the  transition  to a harder
component    {\em appears}  as  a   softening  is  in fact possible
but is  requires  the existence of two  condition,
that  while perfectly possible are, in the absence of
a dynamical  explanation  surprising coincidences:
\begin{enumerate}
\item  Both galactic  and extra--galactic components must have
 a steepening  feature  at approximately
the  same  energy  $E^*$.
The  extragalactic  component  must however   be 
the harder   of the two,   both  below  and above $E^*$.
\item   At the  energy  $E^*$  the 
two   components  must  have  approximately equal  
intensity ($ \phi_{\rm gal}(E^*) \simeq
  \phi_{\rm extra}(E^*)$). 
\end{enumerate}
Let us consider  a situation where
the galactic  component has slopes
$\alpha_{\rm gal}$   and  $\beta_{\rm gal} >  \alpha_{\rm gal}$
 below  and above $E^*$,  while
the extragalactic component  has 
slopes $\alpha_{\rm extra}$  and   $\beta_{\rm extr} > \alpha_{\rm extra}$.
If the  relative normalization  of the two component
is   ``about right'', and if the 4 slopes
are  ordered  as:
$$
\alpha_{\rm extra} < \alpha_{\rm gal}  < \beta_{\rm extra} < \beta_{\rm gal}
$$
then one will observe a  spectrum  that  steepens from
the slope $\alpha_{\rm gal}$ to the slope $\beta_{\rm  extra}$,
with the steepening  marking the transition to the harder  
extra--galactic  component. 
The two slopes $\alpha_{\rm extra}$ 
(extragalactic  component below the transition)
and $\beta_{\rm gal}$
(galactic  component above the transition)
are  not easily measured   because they are ``hidden''
  by  the dominant component.

The discussion above  is just a toy model, but 
in fact it is also an approximation
of what happens in the ``Dip model''.   
At the  transition energy 
(coincident with  ``2nd--Knee''  at   $E^* \simeq  4 \times 10^{17}$~eV)
one observes a   spectrum softening 
from a slope $\alpha \simeq 3.0$  to a  slope  $\simeq 3.3$.
The spectrum below  the  knee
reflects the shape of  a more  abundant 
galactic  component ($\phi_{\rm gal} \propto  E^{-3}$), while the   
harder ($\phi_{\rm extra} \propto E^{-2.4}$) extragalactic  
proton component  is  hidden   below.
Near the transition energy the galactic  component  two things happen:
the galactic  component  steepens sharply  (presumably because
of the ``end'' of the galactic accelerators), while also
the extragalactic component   steepens   because of
the opening up of the pair  production channel for energy loss,
taking  (just below the ``2nd--Knee'')  the  approximate slope  $\sim 3.3$.
The relative  normalization of the  two components 
is  such   that one  observes  a single  
steepening transition (approximated  as the  slope transition
$3 \to 3.3$).
  
This  situation is  possible,   but it is  not  {\em natural},
and  seems to require some ``fine tuning''. 
In fact the same     critical  
observations have already been 
formulated  in the past  discussing the  ``knee'' 
($E_{\rm knee} \sim 3 \times 10^{15}$~eV)
as a possible  transition  between two  different CR populations.
Additional discussion in 
support of the ``Dip model''  can be found 
in\cite{Berezinsky:2005cq} or\cite{Hillas:2006ms}.

\vspace{0.2cm}
In summary,  the  identification of the 
transition energy that  separates galactic  and  extra--galactic cosmic rays 
is a central problem for  high energy astrophysics, 
with wide and important  implications.

Luckily  it is likely that 
the problem will be solved with new  observations
especially those related to the mass composition
and angular  distribution of the cosmic rays.
Essentially all CR models   predict  that  the
two components have very different  mass composition
(with a galactic   component dominated by heavy nuclei, and an
extragalactic one rich  in protons), therefore  the 
measurement of  energy spectra  separately  for different
mass ranges    has the potential to clearly disentangle
the two components.
Similarly, the large scale  angular distribution of   CR
below  and above the transition energy, 
(possibly also  for different mass groups)
can allow to separate an isotropic
extra--galactic component   from a galactic one
that should show the effects of the Milky Way magnetic
confinement.

\subsection{Energy  Balance for Extra--Galactic  CR}
The  average energy  density of  extra--galactic 
cosmic rays   $\rho_{\rm cr}^{\rm extra}$
is  the result of the continuous  injection  of  particles
during cosmological  history, taking into account
the losses due to    energy redshift and other interactions.
Integrating over all   energies  and  considering   only
redshift losses one  has:
\begin{equation}
\rho_{\rm cr}^{\rm extra} = \int dt ~\frac{ {\cal L}_{\rm cr} (t)}{1 + z(t)}
= \int dz ~\frac{  {\cal L}_{\rm cr} (z) }{H(z) \, (1+z)^2}
= \frac{{\cal L}_{\rm cr} (0) }{H_0} ~ f
\end{equation}
where  $ {\cal L}_{\rm cr} (z)$ is the power  density
(in  comoving coordinates)   of the  CR sources
at the epoch $z$  (and  $f \simeq 1--3$ is an adimensional  quantity).
Includig  pair and pion production lossses for protons
require  a more complicated  but straightforward  analysis.

The estimate of the necessary CR power  density is  a very important
constraint    for the possible sources.
This  estimate  is  however  difficult   because
only a  small  portion   of the  extragalactic spectrum  is  visible, 
with  the low energy part  hidden  by the galactic  foreground, 
and therefore the total   energy density is   not directly  measured
and can at best be obtained  with an extrapolation  based on theory.

\begin{figure} [hbt]
\centerline{\epsfig{figure=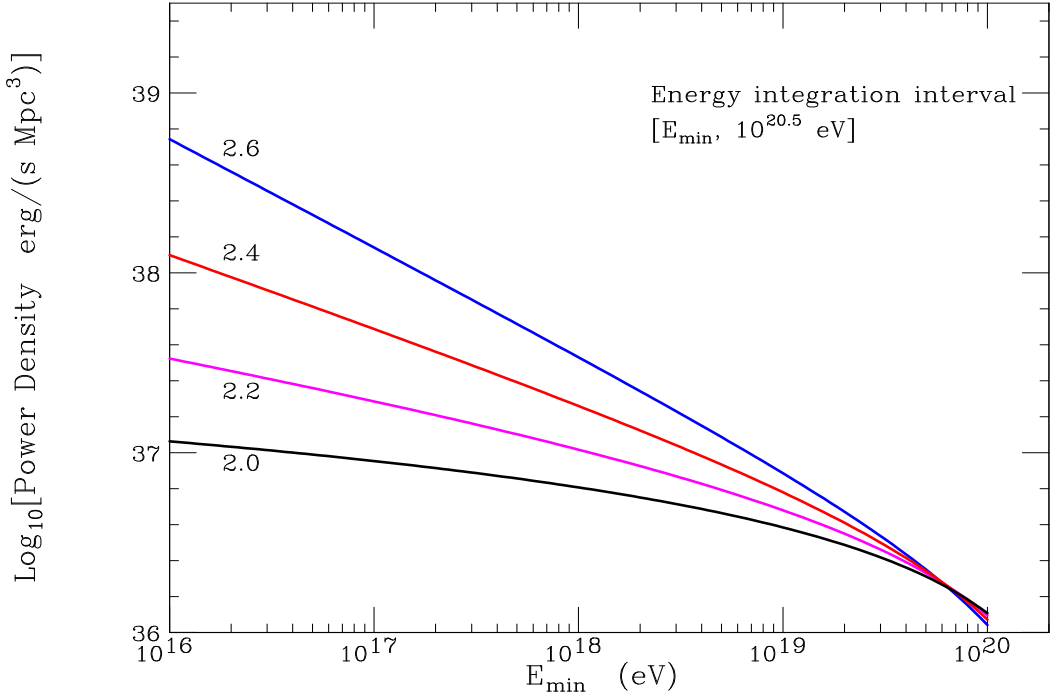,angle=90,width=14cm,height=9.3cm}}
\caption {\footnotesize   
Power density necessary to  produce the extra--galactic  cosmic  rays,
plotted as a function of the minimum energy $E_{\rm min}$  of the
injected particles.
\label{fig:pow0}  }
\end{figure}

It  is  possible  to  estimate  the energy density
(and correspondingly the power  density)   of cosmic  rays
above a minimum  energy $E_{\rm min}$.
Fig.~\ref{fig:pow0}   plots   the  power density  of the
extragalactic proton  sources   that correspond to the 
fluxes given  in fig.~\ref{fig:densa}.
All  these fluxes    reproduce  the  highest energy part 
(above the ``Ankle'') of the
CR spectrum,  but having different  slopes they  have
different  behaviours at lower  energy.  
The required power  density   clearly depends   on the
minimum   energy $E_{\rm min}$, and  is also a    (rapidly varying)
function of the slope $\alpha$.

The power  density needed to generate the CR  above   
$10^{19}$~eV is of order:
$$
 {\cal L}_{cr}^0 [E \ge 10^{19}~{\rm eV}] 
\simeq (3 \div 6) \times 10^{36}~ \frac{{\rm erg}}{{\rm Mpc}^3s} .
$$ 
This
power  estimate depends  only 
weakly on the assumed  shape of the injection spectrum,  however
extrapolating to lower  energy the  power  density 
increases  slowly (logarithmically)  for $\alpha \simeq 2$ 
but  faster ($E_{\rm min}^{\alpha -2}$) for    softer  injection.
In the ``Dip  Model''   where the  CR injection is  relatively soft
($\alpha \aprge 2.4$) and the  minimum energy is 
$E_{\rm min} \aprle 10^{18}$~eV, 
one has a  minimum  power requirement 
${\cal L}_{cr}^0  \aprge  10^{38}$~erg/(Mpc$^3$s), that is  significantly
larger than the previous estimate.
If the   power  law form injection  continues
without  break to  lower energy, the
power requirement    increases  with  decreasing
$E_{\rm min}$  and   eventually becomes extraordinarily large.

What are   the possible candidates for  the sources  of 
the extragalactic CR,  based on  this  power  requirement?
The two leading candidates 
are  Active Galactic Nuclei (AGN)  and Gamma Ray Bursts (GRB).
The   bolometric   power  density of the ensemble of AGN
is uncertain  but  for example  it was estimated  
by Ueda\cite{ueda} multiplying by a  factor of 30
the observed energy density in hard X--ray  ([10,20]~KeV)  band:
\begin{equation}
{\cal L}_{\rm AGN}^{\rm bolometrix}   \simeq 2  \times 10^{40}  
~\left ( \una  \right )
\end{equation}
It is not  unreasonable to  expect that a fraction of
order   1--10\% (or possibly more) of the  released  energy
is  transformed into   relativistic particles, and
therefore AGN  satisy the energy constraint.

Active Galactic  Nuclei are     the central regions of Galaxies 
that  emit  radiation in a  very  broad  energy  range, their source
of  energy is   modeled  as  due  the gravitational  energy
released  in mass  accretion on  supermassive Black Holes 
(with mass as  large as $\sim 10^{9}$--10$^{10}$~$M_\odot$).
It can be  interesting to note that the   power output 
of the ensemble of AGN  can  be checked  against the  observed
mass density in supermassive Black Holes in the universe.
When a  mass element $m$  falls  into  a  Black Hole (BH) of
mass  $M_{\bullet}$, the BH mass  increases  by an
amount
$\Delta M_{\bullet} \simeq (1 - \varepsilon) \, m $, while  the energy 
$\varepsilon  \, m$ is radiated away.
The relation between a  BH mass  and the  energy  radiated 
during its formation  can therefore be  estimated  as:
\begin{equation}
M_{\bullet} = {(1 -\varepsilon) \over \varepsilon } ~ E_{\rm radiated} 
\end{equation}
The   radiation  efficiency  $\varepsilon$ can
be roughly estimated  as the kinetic  energy  per unit mass
 ($G \, M_\bullet /r$)  gained  by a particle  falling into  
the BH potential  down to  a  radius  $r$ equal to a few
 Schwarzschild radii  ($R_S  = 2 \, G \, M_\bullet$), where the 
particle falls is  temporarily arrested  by the collision  with
the  other accreting matter.
For $r \simeq  5 \, R_S$ one  finds  $\varepsilon \sim 0.1$.
The measured density of BH's in the local universe\cite{marconi}
 is
 $\rho_\bullet \simeq 4.6 ~_{-1.4}^{ +1.9}  \times 10^5~M_\odot/{\rm Mpc}^3$,
the corresponding   density in  radiated  energy is
of order $\sim  10^{-4}$~eV/cm$^3$, and if a  fraction of  few percent
of this  energy  is  converted into  relativistic particles,
 this is  sufficient to form the  CR extragalactic   spectrum.

The possibility to  obtain sufficient power
from  Gamma Ray Bursts to fuel the observed  extra--galactic cosmic ray density
is  the object of some  dispute.  
Several authors\cite{Dermer:2007au,Vietri:2003te} have argued in 
favor of  this explanation, 
while according to others
  (see for example\cite{Hillas:2006ms})  the GRB energy output is insufficient.

The long  duration GRB's  have been  solidly associated  with
a  subset of SN explosion.  The SN   blast waves  are considered
as the  source of most of the galactic CR.  
The rate of SN explosions  in the near universe  
is  estimated\cite{Fukugita:2004ee} as:
\begin{equation}
R_{\rm SN}^{\rm observed}  
\simeq  7.6^{+6.4}_{-2.0}   \times  10^{-4} ~ ({\rm Mpc}^{-3} ~{\rm yr}^{-1}) 
\end{equation}
 in good agreement with  theoretical  estimates based
on the   star formation  (and death) rates.
This  SN  rate 
corresponds to the power density  
(in the form of kinetic energy of  the ejecta)
${\cal L}_{\rm SN}^{\rm kin} \simeq 3 \times 10^{40}$~erg/(Mpc$^3$s),
and  presumably to  a  CR  power density 5--10  times smaller.
It is  however assumed  that most of the CR produced around the  spherical
SN shocks  cannot  reach the energy range where the extragalactic  CR 
are  visible.
The  GRB  phenomenon, that is the emission of ultrarelativistic  jets
is associated with a small  fraction  of the GRB, 
estimated  by Woosley and Hower\cite{Woosley:2002zz} as  less than 1\% 
the core collapse SN.    
Estimating a  CR  emission   energy  per GRB of order $10^{51}$~erg 
this can provide   a  CR injection power  density  of order 
$10^{37}$--10$^{38}$~erg/(Mpc$^3$s).
This is sufficient to  generate  
the extragalactic  CR in    models  where the
injection has a  flat spectrum, or where the 
CR emission is  limited to very high  energy,  but it is  likely to
be too small for other models  with  softer injection, and
an emission that   grows  to  $E$ as small as a TeV or less.

\section {Cosmic  Ray Point Sources}
At a sufficiently  high energy   the magnetic deviations of charged  particles
in the propagation from  their sources
should become sufficiently small
so that  it becomes possible to  perform  high  energy cosmic ray  astronomy. 
This simple and very attractive  idea  has  been  for a long time
a ``dream'' for  CR  physicists and, with a  high  degree of  confidence, 
 has recently become a reality thanks to new
data of Auger\cite{Cronin:2007zz,Abraham:2007si}, even if the
result  has   received  only weak or no  confirmation  by
the Yakutsk\cite{Ivanov:2008it} and HiRes\cite{Abbasi:2008md}
detectors, and  the 
interpretation of the result remains   controversial, at the center
of  very lively  discussion.
\begin{figure} [hbt]
\centerline{\epsfig{figure=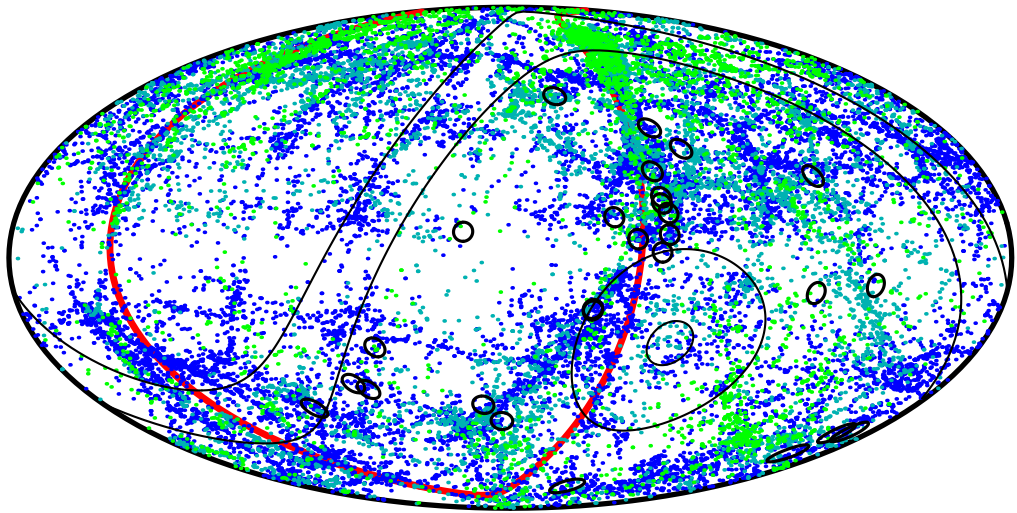,width=14cm}}
\caption {\footnotesize 
Hammer--Aitoff (equal area) projection of the Auger 
(black circles) events with energy
$E \ge 6\times 10^{19}$~eV.  The points are galaxies 
with $z < 0.015$. The thick (red) 
line is the  so  called ``Super--Galactic plane''.
The thin  lines  delimit regions in the sky that have 
received  equal  exposures.
\label{fig:pointa} }
\end{figure}

For quasi--linear  propagation, 
the angular  deviation of a charged particle propagating 
for a distance $d$   in  extragalactic space 
can be estimated  assuming that the  field has a  typical   value $B$, 
and it is  organized in  ``magnetic  domains''  
of linear size $\lambda_B$  where the  field  direction roughly 
parallel.
Summing  the deviations  contributed by
all  magnetic domains  crossed by a  particle
results in  the estimate of the deviation:  
 \begin{equation}
 (\delta \theta)^2   \simeq
 N_{\rm domains} ~\left ( \frac{\lambda_B}{r_L} \right )^2 
 =  \frac{d \; \lambda_B ~ (Z \,e \, B)^2}{E^2} 
 \label{eq:mag_dev}
 \end{equation}
(where $r_L = E/(Z \,e \, B)$ is the Larmor radius). 
Numerically:
\begin{equation}
\delta \theta =  0.53^\circ 
~Z\; B_{\rm nG}  
~\sqrt{(d\; \lambda)}_{\rm Mpc}
~\left ( \frac {10^{20} ~{\rm eV} } {E} \right )
\end{equation}
Astronomy  with CR of  energy $E$ 
 is therefore  possible 
(for an angular resolution of order $\delta \theta$)  only 
within  a sphere  of  radius:
\begin{equation}
R_{\rm imaging}  (E, \delta\theta) =  
\frac{ E^2 \, \delta\theta^2}{(Z \, e \, B)^2 \; \lambda_B }
\label{eq:rimaging}
\end{equation}
The ``CR imaging radius''  is  strongly dependent
($\propto E^2/Z^2$)   
from  the charge and  energy   of the particle  considered, 
 and  shrinks also   rapidly 
 ($\propto (\delta \theta)^2$)  if  one requires  sharper  images.
The  crucial  parameter that  fixes the size of  
the CR imaging sphere  is the  combination $(B^2 \lambda_B)^{-1}$  
that characterizes the  extragalactic  magnetic field. 

At the end of 2007 the Auger collaboration\cite{Cronin:2007zz,Abraham:2007si}
has  published some  potentially very important  results  about 
the  angular distribution  of their  highest energy events.
The Auger group has  shown that 
the  arrival directions of the  26 highest energy events  
($E_{\rm min} \simeq 57 \times 10^{18}$~eV),
and the positions of the  closest   292  Active Galactic  Nuclei 
contained in  the Veron Cetty catalogue,
and in the  detector  field of  view 
with a maximum  redshift $z_{\rm max} = 0.017$  
(corresponding to $d \aprle   71$~Mpc)  are correlated. 
Of the 26 selected events, 20  arrive  within a cone of
aperture  $\psi = 3.2^\circ$ around  one of the AGN  positions. 
The expected  number of  coincidences  for an isotropic
distribution of the  CR  arrival direction
is estimated  as 5.6, and the  significance of the excess,
taking into account the ``statistical cost''
of optimizing the three quantities $\{E_{\rm min}, z_{\rm max}, \psi\}$
is  estimated\footnote{This probability is  estimated
in robust way, ``optimizing''  the   three   parameters
from  only  the  first half of the collected  data, and 
then  computing  
$P$   from the second  half of the data,
 keeping the parameters  fixed} as $P = 1.7 \times 10^{-3}$.

 This  important  result has  attracted  considerable  attention
 an several   comments have  already appeared in the literature.
 The most   straightforward    (or ``naive'') interpretation  of the  results
 is of course that the AGN  themselves are the sources 
 of the highest energy CR, and that  the   angular scale
 of $3.2^\circ$  is    simply  the typical  deviation
 suffered by  the particles  during their propagation from their
 sources.
 This would be an extraordinary  result  that  at the same time
 (a) solves  (or  goes a very long way  toward solving)
 the problem of the UHECR sources; (b)  gives  a  crucial
 measurement  of the  properties of the   extragalactic  magnetic  field;
 and  more controversially (c)  measure the electric charge ($Z\simeq 1$)
 of the UHECR.
 It should be noted that the  angular resolution of the Auger detector
 is  better that  1$^\circ$, and therefore the   3 degrees cone is  not
 a  physical, and not an instrumental effect.  This  deviation is  accumulated
 during the entire   trajectory of the particle, and can be decomposed
 in contributions from the ``source envelope'',  the extragalactic
 propagation and finally  deviations in the
 Milky Way  disk  and  magnetic  halo.
\begin{equation}
(\delta \theta)
 \simeq 
(\delta \theta)_{\rm source}
\oplus
(\delta \theta)_{\rm extra}
\oplus
(\delta \theta)_{\rm MW~halo}
\oplus
(\delta \theta)_{\rm MW~disk}
\end{equation}
Our  knowledge of the  large scale of the  magnetic  field
of our own Galaxy   especially  away from the disk
is   incomplete and  non--precise, however
it is  reasonable  to  expect  that  for the energies
considered   the  integral   of the magnetic field  
 $\int d\vec{\ell}  ~\vec{B}(\ell)$  
over most line of  sights  corresponds
to   deviations  larger than 3$^\circ$ if the charge  much
larger than unity, and therefore one  can use our own galaxy
as  a  spectrometer  to  determine that  the  CR are protons
or at most light nuclei.
Interpreting the dimension of the cone that optimizes
the correlation  with the AGN as the deviation angle
of the particles  one can also
use equation (\ref{eq:mag_dev}).
for a  first  estimate of  $B^2 \, \lambda_B$.
Inserting  the values
$\delta \theta  = 3.1^\circ$, $d \simeq  71$
and  $E = 60 \times 10^{19}$~eV  in (\ref{eq:mag_dev})  one   obtains:
\begin{equation}
Z^2 \, B^2 \; \lambda_B \simeq 0.15  ~({\rm nGauss})^2~{\rm Mpc}
\label{eq:estim1}
\end{equation}
(where $ Z \sim 1$ is the charge of the observed particles).
This  implies  an  imaging radius:
\begin{equation}
R_{\rm imaging}  =  
210~
\left (
\frac{ E}{10^{20}~{\rm eV}} 
\right )^2
~
\left (
\frac{\delta \theta_{\rm image} }{3^\circ} 
\right )^2
~{\rm Mpc}
\label{eq:r_imaginga}
\end{equation}
that is  very encouraging for the perspectives of CR astronomy.

These ``nominal'' interpretation of the Auger data 
has  however   a   number of  significant problem
and it can only considered as    tentative. 
In fact  the   publications of the Auger group  are very careful
in  observe  that   since the AGN  distribution  is  correlated to 
the  distribution of  normal  galaxies.
Moreover, the Auger  data can  can also  
be interpreted\cite{Gorbunov:2008ef,Stanev:2008sd}
assuming  that  the anisotropy  the effect  is  mostly due to the
a   the    contribution   of a single  source,
the  nearby  ($d \simeq  3.5$~Mpc) 
Active Galactic Nucleus  CEN A,  
generating as  much as 
one third of the events above $6 \times 10^{19}$~eV,
with a  larger  angular spread
of order $\delta \theta \simeq 10^\circ$ (or more).
 Substituting the   shorter  distance
and the larger $\delta \theta$ one  obtains  an estimate of 
$B^2 \; \lambda_B$   larger  
by a  factor of order 200:
\begin{equation}
Z^2 \, B^2 \; \lambda_B \simeq 42 ~({\rm nGauss})^2~{\rm Mpc}
\label{eq:estim2}
\end{equation}
and correspondingly a  200  times smaller  imaging radius.
The  discrimination  between  the  two  estimates
(\ref{eq:estim1})  and
(\ref{eq:estim2})  for the  parameter  that describes the 
extragalactic magnetic  field is obviously crucial for the future of
the field.
\begin{figure} [hbt]
\centerline{
\epsfig{figure=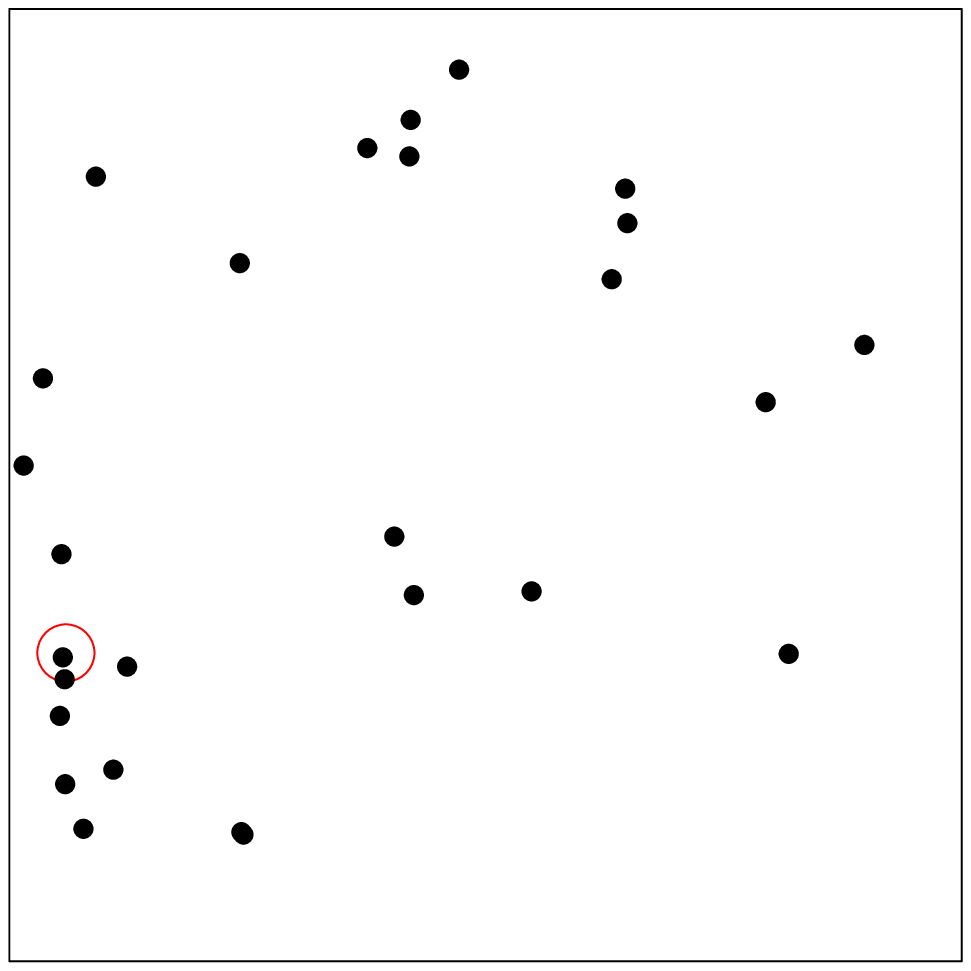,width=6cm}
~~~~
\epsfig{figure=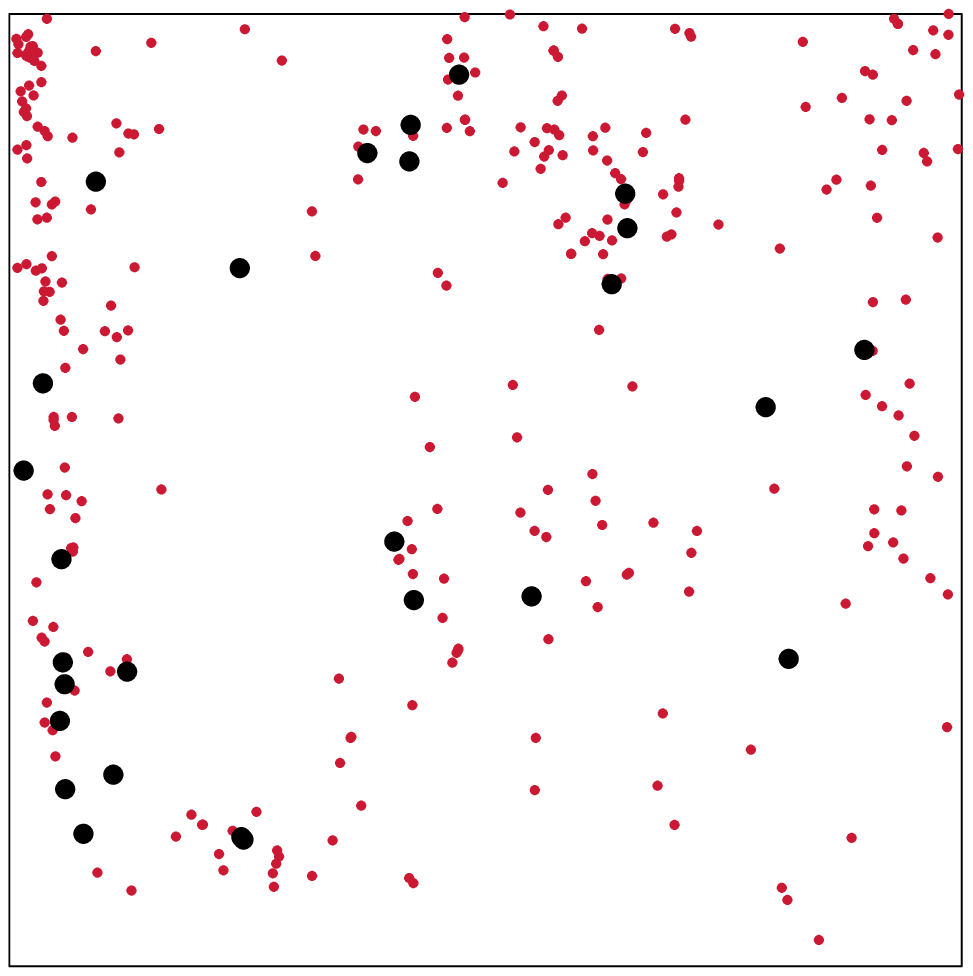,width=6cm}
}
\caption {\footnotesize 
 The left panel  shows the  angular   distribution
of the 26    highest  energy events of  Auger\protect\cite{Cronin:2007zz}.
The  horizontal $x$ axis  corresponds to the right ascension and the 
vertical  $x$ axis to  the declination $\delta$.  The  $y$  coordinates have  
been  ``stretched''   taking into account  the   declination dependence
of the exposure, so that   for  an  isotropic
distribution,   the   points   that correspond
to the measured  directions  should  fill uniformly  the square.
The probability that the measured arrival directions are  distributed
uniformly  is few    percent. 
The  right panel  shows the distribution of  near 
($ z < 0.015$)   AGN in the Veron--Cetty  catalogue.    
The position of CEN A is  also  shown as a red  circle in the left panel.
\label{fig:pointb}  }
\end{figure}

\section{Cosmic Ray Composition and  Hadronic  Interactions}
A  crucial  problem   for  CR  science  is 
the estimate of the energy and mass  of  the detected   particles. 
At high  energy it is only possible to   observe the 
showers  produced  in the atmosphere  by CR particles,  and
to  extract information  about the properties of the  primary particle 
it is obviously necessary to have a  sufficiently 
accurate  description of  the shower  development.
The   key ingredient, and  the dominant source of  systematic  uncertainty
is  the description   of  the properties of hadronic  interactions,
such  as:  cross sections,  final state  multiplicities and inclusive
energy spectra.
At a fundamental level  there  are  few  doubts  that
Quantum--Chromo--Dynamics (QCD)    is a  complete and 
successful theory  of the strong interactions;
however the  fundamental  fields 
that  enter the QCD   Lagrangian  are only quarks and gluons,  
and   perturbative   calculations are  only  possible
for  high momentum   transfer processes  between these fields.
Most of the  observables  that are relevant for shower  development 
cannot be calculated from first principles.
These  quantities (like  cross sections and multiplicities)
must be obtained  from  measurements 
obtained in  accelerator experiments.
The difficulty here is that  the available  data  do not 
entirely cover the relevant  phase space, and therefore
for the interpretation of  CR observations it is 
necessary to  {\em extrapolate}  from the regions  were
observations  are  available,  using   guidance
from the  known  structure of the underlying theory. 

\begin{figure} [hbt]
\centerline{\epsfig{figure=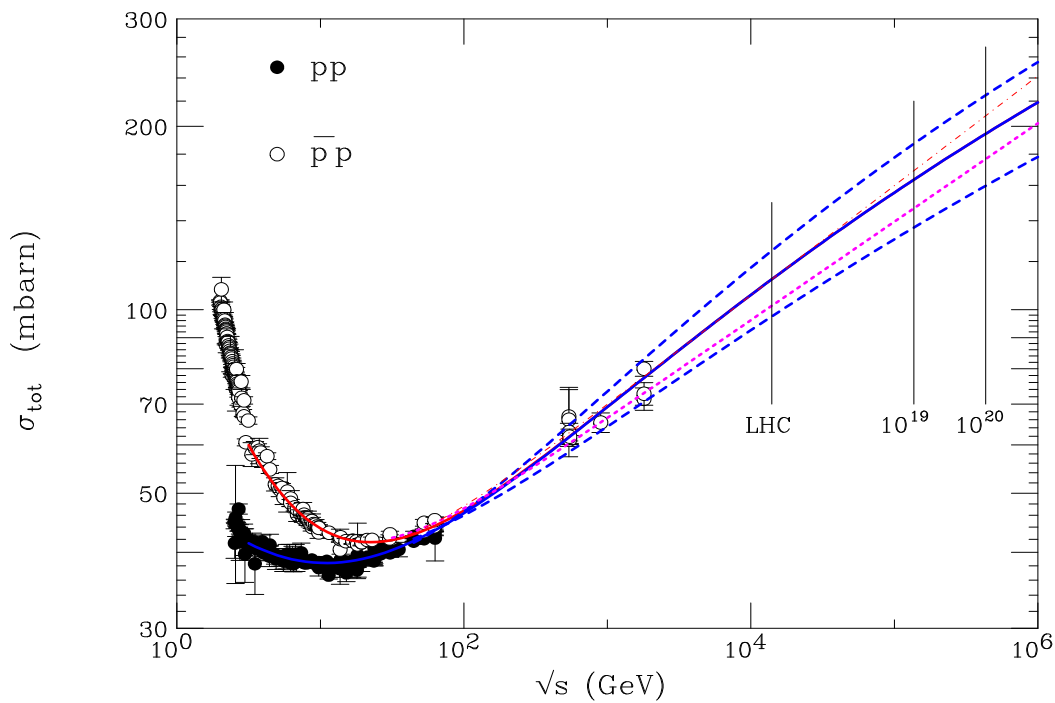,angle=90,width=14cm,height=9.3cm}}
\caption {\footnotesize Total  cross sections
for $pp$ and $\overline{p}p$ scattering. 
The filled (empty) points are  measurements for 
the  $pp$ ($\overline{p}p$)  channel.  The  solid line is  a
is the result of a  fit  in  \protect\cite{Cudell:2002xe}
The dotted  line is  the fit  in \protect\cite{Donnachie:1992ny}
\label{fig:sig_pp} }
\end{figure}

The   incomplete  phase space  coverage is    evident
if we consider   the center of mass  energy of the  first 
CR interaction, that for   the highest energy  particles 
with $E \sim 10^{20}$~eV   is $\sqrt{s} \sim 430$~TeV,
four hundred times  larger than the 
highest energy  where accelerator  data is available
(at the  $p\overline{p}$  Tevatron collider).  This
``energy gap'' will be only  partially filled  by the
eagerly awaited data of the LHC  collider at CERN
with a c.m.  energy $\sqrt{s} = 14$~TeV
(see for  example  a plot of the total $pp$ cross section
in fig.~\ref{fig:sig_pp}).
It should also  be noted that  data  at  high  $\sqrt{s}$ is
available only for $pp$ or   $p\overline{p}$   collisions,
(and not for meson--nucleon  and  hadron--nucleus 
interactions).
Note also  that  for an accurate description of 
the  shower development it is  necessary to have 
a precise  description of the highest energy
secondaries  (in the projectile  fragmentation region),
these particles are  experimentally the most difficult to measure
in collider experiments, and   significant  uncertainties about their
energy spectrum still  exist.  For this reason   even below the knee
($\sqrt{s} \simeq 2.3$~TeV)  uncertaintis in the description of
the hadronic   interactions  are a source of  significant  errors
in the energy  and mass  measurement.

Fortunately, the  uncertainties in  
the description of hadronic interactions 
play only a limited  role in the 
energy determination  for the  highest energy particles, 
where one can observe  showers  using  the  
fluorescence light method  pioneered  by the Fly's Eye  detector.
Relativistic charged  particles in air   excite 
 nitrogen  molecules in the medium.  These excited   molecules
can return  to the  fundamental states  emitting
isotropically fluorescence  photons,   that are  detectable  at  the ground,
with the shower  appearing as a  quasi point--source of light 
moving at the speed of light across the sky.
The photon emission is  proportional to the number of 
charged particles, and thereore 
from the angular distribution of the detected  photons
it is possible to reconstruct  the longitudinal profile
$N_e(X)$ of each detected shower\footnote{
$N_e$ is the number of  charged particles
present at the depth $X$, and
the   column density $X$  can be used as  a  coordinate
along the shower axis}.  
The longitudinal  profile $N_e(X)$ allows  the  model independent
reconstruction of the energy   dissipated in air  by the shower as
 ionization  (and then ultimately heat):
\begin{equation}
E_{\rm ionization} = \int_0^{X_{\rm ground}} dX ~ 
\left \langle \frac{dE}{dX} \right \rangle ~N_e(X)
\label{eq:ionization}
\end{equation}
The total  shower energy can be calculated  adding 
to this dominant  contribution   smaller correction terms
that take into account the energy  
that goes into neutrinos, muons,  and is dissipated in other 
(as  hadronic and electromagnetic  components)  
below the ground:
\begin{equation}
E_{\rm shower} =  
E_{\rm ionization} 
+ E_{\nu} 
+ E_{\mu} 
+ E_{\rm  ground} 
\end{equation}
The correction  terms 
$E_{\nu} $, 
$E_{\mu} $ and
$E_{\rm  ground}$ 
are  mass   and model  dependent,  but fortunately
account  for a small fraction  (of order $\sim 10\%$)
of the primary particle energy, 
and   the energy measurement 
has only a small model dependence.

In other words,  according to  equation (\ref{eq:ionization}),
 the {\em area} subtended by the
longitudinal  profile $N_e(X)$  is a
measurement of the shower energy
that is  independent from the  particle type and
from the modeling of  hadronic  interactions;
 on the other hand the {\em shape}
of the  curve depends on both the  primary particle mass 
and the properties of   hadronic  interactions.

The  shape of the  showers  longitudinal development  is in fact
the best available  method to  estimate the mass  of   UHECR.
It is in fact intuitive that,   for the same   total
energy,  the   showers  generated  by a 
nucleus of  mass $A$    will    develop more rapidly, 
reaching their maximum size   at a shallower depth.
Because of fluctutations  this method cannot  identify
the mass of an individual  shower, but can be used  statistically
to  estimate   the mass  composition at a given  energy.
The   simplest method is to use the average  value
$\langle  X_{\rm max} (E) \rangle$  
of the position of   maximum   size
for   all showers with energy  $E$  as a measurement of the average 
mass of CR  at that  energy.

For a more quantitative  analysis one can 
denote $X_{\rm max}^p (E)$ the    average   position of shower
maximum  for proton  showers  and,  since showers penetration  grows
approximately    logarithmically   with increasing energy,
develop in  first order around energy $E^*$.
Measuring the energy in units of $E^*$ one obtains:
\begin{equation}
X_{\rm max}^p (E)   \simeq
\overline{X}_p(E^*)   +  D_p  ~\ln E
\label{eq:xmax-dev}
\end{equation}
The quantity  $D_p$  gives the 
increase in the average position
of  shower maximum  when the $p$   energy increases
by a factor $e$  and is  known as the proton 
``elongation rate''

To a very good  approximation
the quantity  $X_{\rm max}^A (E)$,  that is
the  average position of maximum
for showers  generated  by a primary nucleus  of mass $A$ 
and  total energy $E$ is
related to the  average  maximum for proton showers 
by the relation:
\begin{equation}
X_{\rm max}^A (E) \simeq
X_{\rm max}^p \left ( \frac{E}{A} \right ) 
\simeq X_{\rm max}^p (E) - D_p \;  \ln A  
\label{eq:xmax-a}
\end{equation}
In  (\ref{eq:xmax-a}) 
the first equation   is  clearly correct 
for  a naive superposition model
(where  the shower generated by  a nucleus    containing $A$  nucleons
is  described as  the superposition  
of $A$  nucleon showers    each having  energy $E/A$), but 
in fact it  has a much  more general  validity\cite{Engel:1998pw},
since it is  based on    the relations between the 
cross  sections  for  nuclei  and  nucleons 
in the  framework of the Glauber model\cite{Glauber};
the second  equality in (\ref{eq:xmax-a}) follows  from 
(\ref{eq:xmax-dev}).

More in general, if  the CR  of  energy $E$  are  
a  mixture  of nuclei  with different  mass   $A$,
 the average  $\langle X_{\rm max} \rangle$   
can  be  expressed as:
\begin{equation}
\left \langle 
X_{\rm max} (E) 
\right \rangle
\simeq 
X_{\rm max}^p (E) 
- D_p (E) \; \left \langle \ln A\right \rangle_E
\simeq \overline{X}_p + D_p \; \left [ \ln E - \langle \ln A \rangle_E \right ]
\label{eq:loga}
\end{equation}
where 
$\left \langle  \ln A  \right \rangle_E$  the  average 
value  the  logarithm of $A$  for  particles of energy $E$.
One can  then use equation 
(\ref{eq:loga}) to   estimate  the mass  composition of UHECR
comparing the  measured  $\langle  X_{\rm max}\rangle$  with the
value expected  theoretically for protons:
\begin{equation}
\langle \ln A \rangle_E = 
\frac{\langle X_{\rm max} (E) \rangle - X_p (E)}{D_p}
\end{equation}
Similarly  comparing the experimental  elongation rate
$D_{\rm exp} =  d\langle X_{\rm max} (E)\rangle/d\ln E$ with   the
elongation rate  for protons  (or nuclei)  $D_p$, 
 one  can estimate the rate of
variation of the composition with energy:
\begin{equation}
\frac{d \langle \ln A \rangle_E}{d \ln E}
= 1 - \frac{D_{\rm exp}}{D_p}
\end{equation}

\begin{figure} [hbt]
\centerline{\epsfig{figure=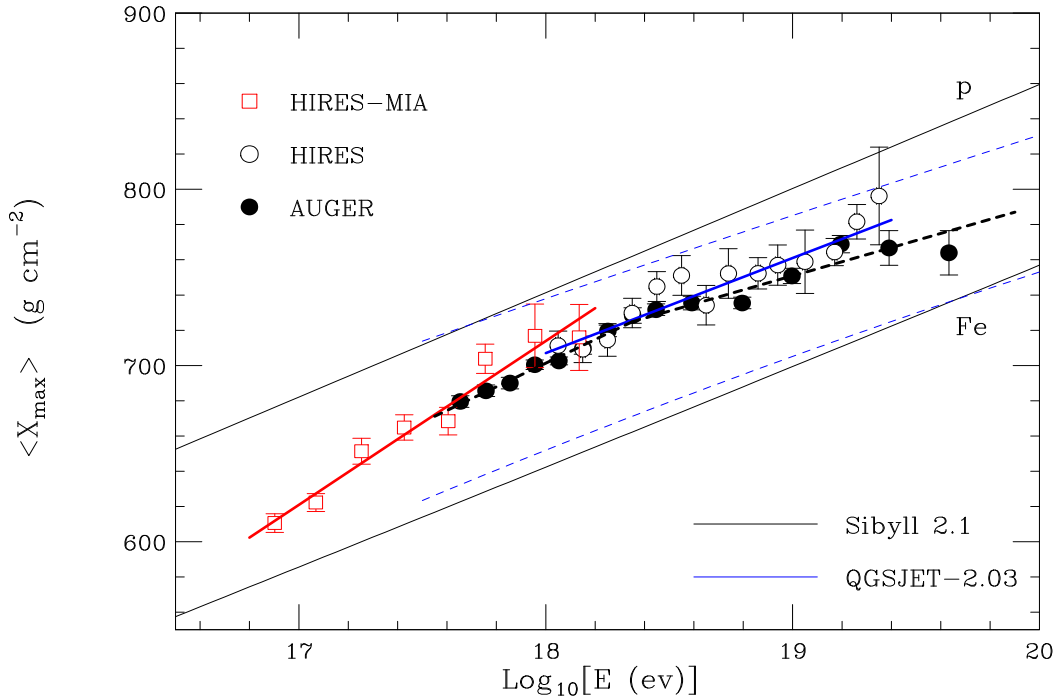,angle=90,width=14cm,height=9.3cm}}
\caption{\footnotesize  
Measurements of  $\langle X_{\rm max} \rangle$
obtained  by the HiRes-MIA  \protect\cite{hires-mia-comp},
HiRes  \protect\cite{hires_comp}  and 
Auger\protect\cite{auger_merida_composition} experiments.
The lines are   montecarlo calculations  for protons ($p$) and
iron nuclei (Fe)  performed using the  SIBYLL \protect\cite{sibyll0}
and QGSJET \protect\cite{qgsjet} interaction models.
\label{fig:xmax}   }
\end{figure}

An illustration of the present  situation 
of the  mass composition of  UHECR  is   shown in
fig.~\ref{fig:xmax}.  
The points in the figure show the 
average  $X_{\rm max}$  for  UHE  showers   obtained  by the 
HiRes--MIA\cite{hires-mia-comp}, 
HiRes\cite{hires_comp}  and Auger\cite{auger_merida_composition}
experiments.  The thick lines are   fit to the data
points  obtained by the experimental  groups.
while the  thin lines   in  the figure are theoretical  predictions
calculated assuming  a pure composition of only protons
or  only iron nuclei
based on  the hadronic models  of 
Sibyll\cite{sibyll0} and  QGSJET\cite{qgsjet}.
At face  value  these results  seems to  indicate
that  the UHECR  are not  pure   protons  nor
pure iron nuclei,  but   are  a  combination
(with  approximately  equal  weight) of the two species,
or  are  mostly nuclei of intermediate mass.
This  conclusion however  relies on the assumptions  that 
the models  used for the   description of the  hadronic
interaction are at least approximately correct.

The composition of the CR  has  very important  consequences
for the properties of their acceleration sites,   and therefore
it is  of  great importance  to gain a more robust 
understanding   of the relevant properties
of  hadronic  interactions, and a better estimate of the  systematic
uncertainties.
It  is clear for example  that  if the cross  section is larger
or if the energy spectra of the final state particles  are softer 
than what is assumed in the theoretical prediction,  the real showers
develop faster than  the  simulated ones, and the composition
interpretation is  biased  in the direction of a too  heavy composition.

In the discussion of the  previous paragraph,  information
about elementary particle physics (obtained  from accelerator
experiments) is used to  measure  the mass  composition 
of the   highest energy CR, and then  infer the 
structure and properties of their astrophysical  sources.
It is interesting that this 
``logical flow''  can be in principle be  ``inverted''.
In fact  there are  (at least in principle)
methods  to measure   the composition of CR  that are independent
from  the detailed properties of their showers.  These methods,
as  discussed  above, are based on essentially two ideas:
(i)  the ``cosmic  magnetic  spectrometer'', and (ii) the imprints
of energy losses  on a smooth injection sprectrum.
If the ``nominal interpretation''  of
the Auger anisotropy analysis\cite{Cronin:2007zz}
is correct  and  the   deviations of  extragalactic  particles
with  $E \sim   6 \times 10^{19}$~eV 
is as  small as $3^\circ$  their electric 
charge is in fact strongly costrained,
by our (however still poor)  knowldge of the galactic  magnetic fields.
Similarly,  if the ``Dip Model'' of 
 Berezinsky and collaborators\cite{Berezinsky:2005cq} is correct
and the structure of the ankle  corresponds  to the effects
of $e^+e^-$ pairs  production  in $p\gamma$ interactions, 
then  the identity of the UHECR particles is  determined.
In this  case, the requirement of   consistency  for the
penetration   of the  corresponding showers   can be   used  
to obtain  information about  the properties
of hadronic  interactions.

\section{Outlook}
The  observation and understanding of the  ``High Energy Universe''
is one of  the important  frontiers of  fundamental  Science.
Our universe  contains   astrophysical  objects  that are  capable
to accelerate particles to    ultrarelativistic  energies
as large as $\sim 10^{20}$~eV.  The  nature and 
structure of these  cosmic  accelerators and the physical
mechanism  that are operating in them  are  beginning to be clarified,
even if large uncertainties  still exist.  In  fact,   several  classes
of these cosmic  accelerators  are  now  known  to exist. 
There is  good evidence that hadronic  particles  (protons and nuclei)
are accelerated in  SuperNova Remnants  and  Active Galactic Nuclei,
while electrons/positrons  are also  efficiently accelerated
in  several other  environments  like  pulsars and  pulsars wind nebulae.
Gamma  Ray Bursts are also  clearly sites of  particle acceleration.
These  extraordinary   environments are  very important
astrophysical  laboratories to test  our  physical  theories.

The  acceleration of  charged particles is  unavoidably connected  to the
radiation of  photons  and  (in the case of   hadronic  particles) 
to the emission of  neutrinos. Moreover  many of these accelerators
are connected to the violent motion  of large masses and to the
possible emission of gravitational  waves.  Therefore the
complete study of these objects must rely  on the combined  use
of different  ``messengers''  (photons  in  a  very broad
range of  wavelength, cosmic  rays,   neutrinos  and gravitational  waves).
We are now  witnessing the opening of new ``astronomies'' 
with  cosmic ray particles  (at sufficiently high energy  to  reduce
the bending  by magnetic  fields), neutrinos  and gravitational waves.
The  nearby active galactic  nucleus Cen A is  very likely
the first  astrophysical  objects  imaged with  charged particles.
The  development of multi--messenger  astronomy 
is a  revolution  comparable to the  use  of the telescope
for  astronomical  observations started  by Galileo   in 1609,
nearly  exactly four centuries ago, and  is certainly going 
to produce  exciting  results. 

\section{Acknowledgements}
I  would like to acknowledge discussions with Ralph Engel, 
Tom Gaisser, Maurizio Lusignoli,
Todor Stanev and Silvia Vernetto.  Many  thanks  to Milla  Baldo Ceolin
for the organization of another  very fruitful workshop.

\end{document}